\documentclass[journal]{IEEEtran}

\usepackage{xcolor,soul,framed} %,caption

\colorlet{shadecolor}{yellow}
\usepackage[pdftex]{graphicx}
\graphicspath{{../pdf/}{../jpeg/}}
\DeclareGraphicsExtensions{.pdf,.jpeg,.png}

\usepackage[cmex10]{amsmath}
\usepackage{array}
\usepackage{mdwmath}
\usepackage{mdwtab}
\usepackage{eqparbox}
\usepackage{url}
\usepackage{commath}
\usepackage{amssymb}
\usepackage{comment}
\usepackage{amsmath}
\usepackage{cite}
\usepackage{placeins}
\usepackage{float}
\usepackage{makecell}
\usepackage{multirow}
\usepackage{multicol}
\usepackage{graphics}
\usepackage{subcaption}
\usepackage{graphicx}
\usepackage{xcolor} 
\usepackage{soul}
\usepackage{tikz}
\newcommand\copyrighttext{%
  \footnotesize This work has been submitted to the IEEE for possible publication. Copyright may be transferred without notice, after which this version may no longer be accessible.}
\newcommand\copyrightnotice{%
\begin{tikzpicture}[remember picture,overlay]
\node[anchor=south,yshift=10pt] at (current page.south) {\fbox{\parbox{\dimexpr\textwidth-\fboxsep-\fboxrule\relax}{\copyrighttext}}};
\end{tikzpicture}%
}

\begin{document}
\bstctlcite{IEEEexample:BSTcontrol}
    \title{Synchronous Inductor Switched Energy Extraction Circuits for Triboelectric Nanogenerator}
  \author{Madhav~Pathak,~\IEEEmembership{Student Member,~IEEE,}
      and~Ratnesh~Kumar,~\IEEEmembership{Fellow,~IEEE}% <-this % stops a space

 %\thanks{Our conference paper \cite{pathak2018modeling} studied the EECs analyzed here in a rudimentary way and also without any experimental validation.}
  
  \thanks{This work was supported in part by U.S. National Science Foundation under grants NSF-CCF-1331390, NSF-ECCS-1509420, NSF-PFI-1602089, and NSF-CSSI-2004766.}

  \thanks{The authors are with the Department of Electrical and Computer Engineering, Iowa State University, Ames, IA, 50011 USA (e-mail: mpathak@iastate.edu; rkumar@iastate.edu).}
  \vspace*{-.3in}
 }  
 %\pagenumbering{gobble}

%\markboth{IEEE TRANSACTIONS ON POWER ELECTRONICS, VOL.~xx, NO.~xx, MONTH~YEAR
%}{Roberg \MakeLowercase{\textit{et al.}}: High-Efficiency Diode and Transistor Rectifiers}

\maketitle
\copyrightnotice
\begin{abstract}
Triboelectric nanogenerator (TENG), a class of mechanical to electrical energy transducers, has emerged as a promising solution to self-power Internet of Things (IoT) sensors, wearable electronics, etc. The use of synchronous switched energy extraction circuits (EECs) as an interface between TENG and battery load can deliver multi-fold energy gain over simple minded Full Wave Rectification (FWR). This paper presents a detailed analysis of Parallel and Series Synchronous Switched Harvesting on Inductor (P-SSHI and S-SSHI) EECs to derive the energy delivered to the battery load and compare it with the standard FWR (a 3rd circuit) in a common analytical framework, under both realistic conditions, and also ideal conditions. 
Further, the optimal value of battery load to maximize output and upper bound beyond which charging is not feasible are derived for all the three considered circuits. These closed-form results derived with general TENG electrical parameters and first-order circuit non-idealities shed light on the physics of the modeling and guide the choice and design of EECs for any given TENG. The derived analytical results are verified against PSpice based simulation results as well as the experimentally measured values.
\end{abstract}
% === KEYWORDS ==============

% \fi
\IEEEpeerreviewmaketitle

\vspace*{-.15in}
\section{Introduction: Motivation and objectives} 
\vspace*{-.05in}
\IEEEPARstart{O}{ne} of the major challenges faced by today's ever-expanding field of mobile electronics, Internet of Things (IoT), implantables, and wearable electronics is limited onboard battery lifetime. Placing a high capacity battery presents the demerits of increased size, weight, maintenance, and cost. Hence, recent years have seen a tremendous interest in integrating a miniaturized energy harvester to increase the battery life by scavenging the ambient energy\cite{shaikh2016energy}. Many potential sources of energy are available in the environment, such as mechanical, solar, thermal, and chemical\cite{tan2010optimized,buchli2014towards,shi2014novel,qiao2011remote}. Among these, tapping the mechanical energy in the form of ambient vibration, human body motion, airflow, ocean currents, etc., is almost always possible. Hence, mechanical to electrical energy transducers have received significant attention both academically and commercially\cite{siang2018review}.

Recently, a new class of mechanical to electrical energy transducers termed triboelectric nanogenerators (TENGs) based on the conjugation of friction-induced electrification with electrostatic induction as its working principle have been developed\cite{fan2012flexible}. A major advantage of TENG is almost all materials have triboelectric properties, offering the flexibility of design, fabrication, and operation mode to suit the application\cite{wang2016triboelectric}. TENG based on low-cost paper, stretchable interwoven threads, printed circuit board (PCB), and environmentally degradable materials are some of the many examples, demonstrating the versatility of TENG \cite{zhang2017penciling,seung2015nanopatterned,han2015high,pan2018fully}. TENGs have been employed in terms of application, ranging from harnessing heartbeat vibration for implantables to ocean wave motions for large scale energy harvesting \cite{zheng2016vivo, wang2017toward}.

There remain engineering and technical challenges, e.g., integrating TENG with electronic devices for practical applications, optimal design of transducer, and energy extraction: A critical issue arises from the fact that TENG's output harnessing energy from environment with fluctuating movement is non-D.C. and random. Hence, it needs to be ``rectified" to a stable positive form to be able to charge an on-board battery or capacitor. Also, TENGs are capacitive in nature with high inherent impedance, which might be fixed or non-linearly varying based on the mode of TENG\cite{zi2015standards}. Hence, directly charging battery or capacitor with TENG will in general, lead to a poor charging efficiency due to a severe impedance mismatch. Thus, an interface circuit referred to as Energy Extraction Circuit (EEC) is required between a TENG and energy storage (battery or capacitor). 

Full Wave Rectifier (FWR) is one of the simplest EEC in terms of implementation \cite{niu2015optimization,zi2016effective}. However, it has been demonstrated in the literature that higher per-cycle energy output can be achieved by more complex EEC architectures such as switched circuits\cite{cheng2019power}. Our previous work introduced synchronous switched inductor circuits as EECs for TENG, which involve switching at the TENG voltage output extrema to obtain manifold gain over the FWR output. Wherein, Parallel Synchronous Switched Harvesting on Inductor (P-SSHI), and Series Synchronous Switched Harvesting on Inductor (S-SSHI) EECs for TENG transducers were introduced for the first time to best of our knowledge \cite{pathak2018modeling}. This work extends it by providing a full mathematical derivation, a complete simulation and includes additional results on hardware implementation and validation with the following contributions:
\begin{itemize}
  \item The closed-form equations for the {\em per-cycle energy output} ($E_{cycle}$) of both the P-SSHI and S-SSHI circuits at a given battery load are derived for the first time. In addition, the {\em upper limit on battery load, optimal battery load, and corresponding maximum $E_{cycle}$ for both the SSHI circuits} are also derived. Effect of TENG parameters (determined by construction and operation of the given TENG) and circuit non-idealities on the $E_{cycle}$ have been captured in our models to guide the design of these EECs.
  \item Incorporation of the following critical non-idealities in the analytical model to study their adverse effect on the output energy ($E_{cycle}$) and also to provide a more accurate model for both the SSHI circuits:
  \begin{itemize}
   \item Leakage charge through any off switch over a cycle and also due to other sources such as ringing transients of oscillators at switch toggle times;
  \item Limited quality factor of the resonator circuits determined by parasitic resistance of inductor and the on-state resistance of switches;
   \item On-state voltage drop of diodes.
  \end{itemize}

  \item Experimental implementation of FWR and both the SSHI circuits using self-propelled electronic switches with a simple control circuit comprising just one active component (a comparator).
  \item Validation of the presented analytical models by way of close match with those experimentally measured and PSpice simulated $E_{cycle}$ at different battery loads.
  \item Comparison of P-SSHI, S-SSHI, and the standard FWR circuit in a common framework to bring forward their respective pros and cons.
\end{itemize}

It is important to note that while these EECs have been explored previously for piezoelectric transducers \cite{dicken2012power,singh2018self,singh2015broadband}, their models and hence analysis are entirely different in the case of triboelectric transducers as discussed ahead, mainly owing to the fact that while the piezoelectric capacitor appears in parallel to the load, the triboelectric capacitor appears in series. Also, while the capacitance remains constant in the former, it can be time-varying in the latter.

\vspace*{-.15in}
\subsection{Related Works}
\vspace*{-.05in}
A variety of EEC architectures have been presented in literature\cite{cheng2019power}. A Half Wave Rectifier with a diode parallel to TENG was used to increase the energy output compared to FWR at an optimized load battery voltage. Further improvement was achieved by employing Bennet's voltage doubler circuit \cite{ghaffarinejad2018conditioning}. A transformer with optimal turns ratio can be used for impedance matching; however, the transformer is efficient only in the designed narrow frequency band \cite{pu2016efficient}. 
Synchronous switched EECs include the approach by \cite{niu2015universal} and \cite{harmon2020self} of using a logic controlled switch to extract energy from a capacitor, being charged by a FWR circuit at an optimal instant, namely at an optimal capacitor voltage determined from TENG parameters. TENGs are characterized by high voltage and low transferred charge (or current). To address this condition, \cite{zi2017inductor} presented a mechanically triggered extrema-seeking series-parallel switched capacitor scheme that reduces output voltage and increases charge in corresponding proportion. Also, a mechanically triggered parallel switch to short the TENG at the extrema was used in \cite{zi2016effective} to improve the efficiency and increase saturation voltage for capacitor load. Another approach is using a synchronous serial switch to extract energy from the TENG capacitor at the extrema. This extracted energy can then be transferred to the DC load with high efficiency by using a buck converter as shown by \cite{xi2017universal} or using coupled inductors (flyback converter) as demonstrated by \cite{cheng2017high} and \cite{perez2016triboelectric}. Another example of a switched EEC is the use of Maximum Power Point Tracking (MPPT) for TENG executed using dual-input buck converter over a compact integrated circuit (IC) \cite{8876711}.

\cite{kara202070} recently described the P-SSHI circuit's implementation as an IC. Our work provides for the first time, the analytical models for P-SSHI and S-SSHI circuits, that can help guide the EEC design, such as determining optimal load voltage to maximize energy extraction. \cite{li2019sshi} designed the P-SSHI circuit for a TENG paired with a parallel capacitor that provides an additional degree of controlling the variation between max to min capacitance ratio. In contrast, our work deals with both P-SSHI and S-SSHI circuits in a fully generalized setting and also provides closed-form $E_{cycle}$ equations. \cite{xu2019boost} demonstrated P-SSHI and S-SSHI circuits for a resistive load, one that does not need rectification, using mechanical switches. In contrast, our work studies both the SSHI circuits for varying battery loads commonly encountered in the practical IoT applications; uses convenient automated electronic switches for physical implementation, and demonstrates the desired match between the experimental, simulated, and the derived analytical $E_{cycle}$ results.

\vspace*{-.15in}
\subsection{Comments about Analysis Methodology}
\vspace*{-.05in}
For the different EECs (FWR, P-SSHI, S-SSHI), we derive the per-cycle energy delivered to a battery load rather the instantaneous output power since the vibration frequency can vary, affecting the power but not the per-cycle energy. A battery load is selected for analysis (as opposed to a resistive or capacitive load) since most frequently, there is an onboard battery for IoT, wearables and, other mobile electronic devices, and the harvester's role is to supplement their energy and extend the lifetime. The use of battery load as opposed to resistive load introduces additional complexity of rectification that our EECs also incorporate. The analytical derivations additionally include the computation of the optimal value of the battery load to maximize output and upper bound beyond which the charging becomes infeasible, providing additional insights into the 3 EECs.

Our analytical derivations and experimental verification are provided for the most generalized case: ``Contact-Separation" mode TENGs, in which the internal capacitance varies with time, so that the ratio of maximum to minimum TENG capacitances $\beta\geq 1$. This general class also includes Lateral Sliding and Single Electrode Contact mode TENGs. In contrast, the Sliding Freestanding and Contact Freestanding mode TENGs, having fixed internal capacitance ($\beta=1$)\cite{zi2015standards}, are a special case and can be derived by equating the maximum and minimum capacitances ($C_{T,max}$ and $C_{T,min}$) of our framework. 

Both the SSHI circuits operate with an external inductor, connected briefly at the TENG operation extremes to form a $LC$ oscillator along with the TENG capacitor. For accurate modeling, we carefully consider the quality factor of this oscillator emanating from the non-idealities: parasitic series resistance of inductor ($R_L$) as well as the on-state resistance ($R_{on}$) of the switch. Additionally, the leakage charge ($Q_L$) through the off switches and during the oscillator ringing transients over an operation cycle is also included in our mathematical modeling. Further, the diode's on-state voltage drop ($V_D$) is the third analytically modeled non-ideality in our setup.

\vspace*{-.15in}
\section{TENG Circuit Model}\label{TENG_Circuit_Model}
\begin{figure}[htbp]
\vspace*{-.15in}
  \begin{center}
  \includegraphics[width=1\linewidth]{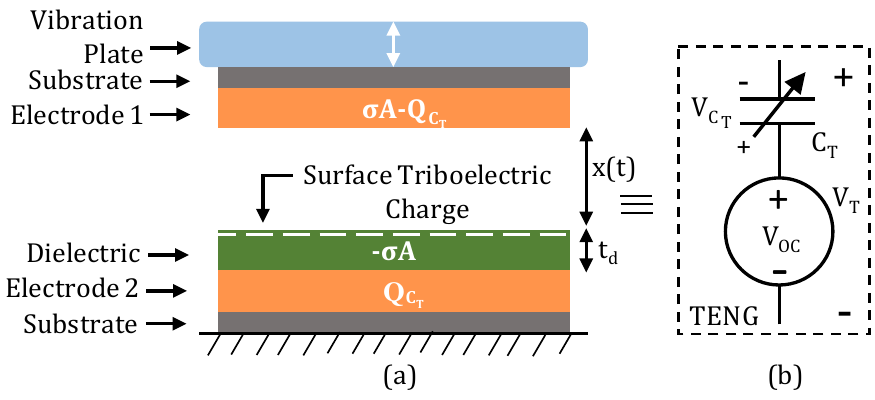}
  \vspace{-.25in}
  \caption{(a) TENG cross section; (b) Equivalent circuit model}\label{model}
  \end{center}
\vspace*{-.25in}
\end{figure}

\vspace*{-.05in}
Here we summarize a 1st-order lumped circuit model for Contact-Separation mode TENG  \cite{niu_zhou_wang_liu_lin_bando_wang_2014}.  Fig.~\ref{model}(a) shows the structure of TENG used for experimental validation of the circuit analysis in this study. 
It consists of a parallel plate variable capacitor with horizontal Aluminum plate electrodes (in orange). The top electrode attached to a vibrating platform can move vertically, making contact and separation with the fixed bottom electrode with a dielectric (Teflon) layer laid over its upper face. The plates have an area $A$ and develop a charge with density $\sigma$ when the two triboelectrics, Aluminum on top vs. Teflon on the bottom, come into repeated contact. Due to Aluminium vs. Teflon's triboelectric properties, the top plate is positively charged while the bottom plate is negatively charged. It can be reasonably assumed that the triboelectric charges (-$\sigma A$) generated on the bottom plate's dielectric layer are evenly distributed, close to the upper surface. When connected to a circuit, the two plates of the TENG will draw/supply charge from/to the circuit due to the electric field of the triboelectric charges. Letting $Q_{C_T}(t)$ denote the corresponding charge drawn on the bottom electrode of the TENG at time $t$, the total charge on the bottom structure at time $t$ is $(Q_{C_T}(t)-\sigma A)$, and by charge conservation, the top plate will have $(\sigma A-Q_C(t))$ charge at time $t$.

Using Gauss law of electrostatics, under infinite parallel plate approximation (no fringing field), the electric fields in the dielectric and air regions of the structure are given by:
\vspace*{-.05in}
\begin{equation}\label{E}
    E_{d} = \frac{Q_{C_T}(t)}{\epsilon_0\epsilon_{d}A}; \mbox{  } E_{a} = \frac{Q_{C_T}(t)-\sigma A}{\epsilon_{0}A},
\vspace*{-.05in}
\end{equation}
where $\epsilon_0$ is the electrical permittivity of air, and $\epsilon_{d}$ is the relative dielectric constant of the Teflon dielectric layer. Denoting the thicknesses of the dielectric and air-gap as $t_{d},x(t)$ respectively, the voltage across the two electrodes of TENG can be calculated using the Poisson's equation, as below:
\vspace*{-.05in}
\begin{equation}\label{V_integ}
\begin{split}
V_T(t) & = -\int_{0}^{t_{d}} E_{d} dx-\int_{t_{d}}^{x(t)+t_{d}} E_{a} dx
\end{split}
\vspace*{-.05in}
\end{equation}
Inserting (\ref{E}) in (\ref{V_integ}), the voltage across TENG can be written as:
\vspace*{-.05in}
\begin{equation}\label{V}
V_T(t) = \frac{\sigma x(t)}{\epsilon_{0}}-\frac{Q_{C_T}(t)(x(t)+\frac{t_{d}}{\epsilon_{d}})}{\epsilon_{0}A}
\vspace*{-.05in}
\end{equation}

From (\ref{V}), we can arrive at the lumped element circuit model of the 
``contact-separation mode TENG" shown in Fig.~\ref{model}(b):
\vspace*{-.05in}
\begin{eqnarray}
V_T(t)&\!\!\!\!=\!\!\!\!&V_{oc}(t)-\frac{Q_{C_T}(t)}{C_T(t)}=V_{oc}(t)-V_{C_T}(t),\mbox{ with}\label{VTENG}\\
\vspace*{-.05in}
V_{oc}(t)&\!\!\!\!:=\!\!\!\!&\frac{\sigma x(t)}{\epsilon_{0}}
\label{Voc}\\
\vspace*{-.05in}
C_T(t)&\!\!\!\!:=\!\!\!\! & \frac{\epsilon_{0} A}{x(t)+d_{eff}}; \mbox{  } d_{eff}:=  \frac{t_{d}}{\epsilon_{d}}.\label{CTENG}
\vspace*{-.05in}
\end{eqnarray}
It is clear from (\ref{VTENG})-(\ref{CTENG}) that TENG capacitance $C_T(t)$ and voltage $V(t)$ change as the air-gap $x(t)$ varies. 

For the analysis purposes, we take $x(t)$ to be varying with period $T$, with minimum value $0$ (when the top plate is fully down) and maximum value $x_{max}$ (when the top plate is fully up). Then at each integer multiple of the period ($t=nT$), the two plates are in contact ($x(t)=0$), whereas at the corresponding half-period later ($t=(n+\frac{1}{2})T$), the two plates are maximally separated ($x(t)=x_{max}$). Accordingly, there are two extreme states of the TENG operation:
\begin{itemize}
    \item State I: $t=nT, x(t)=0, C_{T}(t)=C_{T,max}:=\frac{\epsilon_{0} A}{d_{eff}},V_{oc}(t)=0$;
    \vspace{3pt}
    \item State II: $t=(n+\frac{1}{2})T, x(t)=x_{max},C_{T}(t)=C_{T,min}:=\frac{\epsilon_{0} A}{x_{nax}+d_{eff}}, V_{oc}(t)=V_{oc,max}:=\frac{\sigma x_{max}}{\epsilon_{0}}$.
    \vspace{3pt}
\end{itemize}
The voltage and capacitor values at these extreme states, namely, maximum open circuit voltage ($V_{oc,max}$), minimum TENG capacitance ($C_{T,min}$), and maximum TENG capacitance ($C_{T,max}$) constitute the TENG parameters that influence the energy output of the EECs. A convenient useful term in subsequent analysis is the ratio of the capacitances at the two extremes: $\beta := \frac{C_{T,max}}{C_{T,min}}$.

\vspace*{-.15in}
 \section{Analysis of Energy Extraction Circuits}\label{EEC}
\vspace*{-.05in}
For each of the three circuits, the general approach to derive the per-cycle energy delivered to the battery load ($E_{cycle}$) is to track the TENG voltage at States I and II of the circuit operation. Since the capacitances at States I and II ($C_{T,min}$ and $C_{T,max}$) are known, the corresponding charge on the TENG capacitor ($Q_{C_T}$) can then be found using the TENG equation ({\ref{VTENG}}). By law of conservation of charge, the change in the TENG capacitor charge over a cycle determines the charge flowing through the battery load and hence the delivered per-cycle energy is simply a value scaled by the battery value. Thus, the following analysis and results are purely dependent on the two extrema states mentioned above and independent of the TENG motion path to get to those two extrema. In other words, for a given extraction circuit, $E_{cycle}$ is dependent on the separation at maxima ($x_{max}$ and the corresponding $V_{oc,max}$) and not on the path, $x(t)$. 

\vspace*{-.15in}
\subsection{Full wave rectifier}\label{FWR_Section}
\begin{figure}[htbp]
\vspace*{-.1in}
  \begin{center}
  \includegraphics[width=.87\linewidth]{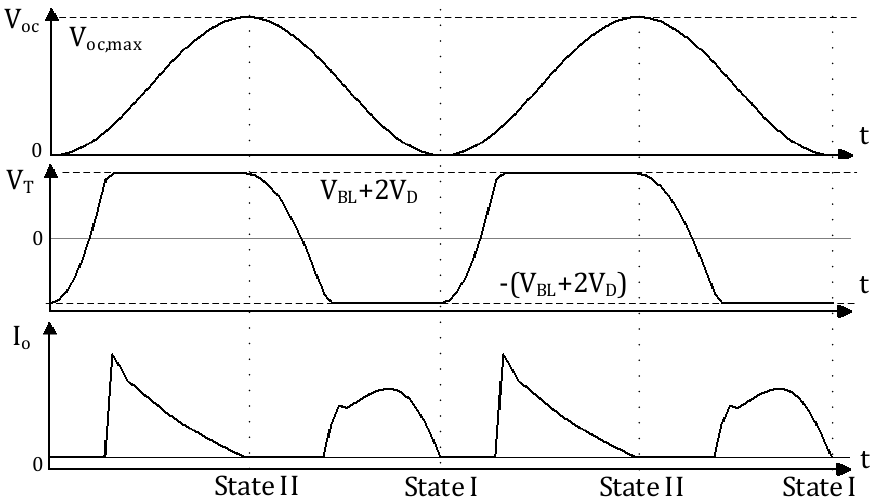}
  \vspace{-.1in}
  \caption{Typical open circuit voltage: $V_{oc}(t)$, FWR circuit TENG voltage: $V_T(t)$ and output current: $I_o(t)$.}\label{FWR_plot}
  \end{center}
  \vspace*{-.35in}
\end{figure}

\vspace*{-.05in}
For a FWR, its battery charging process can be understood by following the voltage waveform shown in Fig.~\ref{FWR_plot}. 
\begin{figure}[htbp]
\vspace*{-.175in}
  \begin{center}
  \includegraphics[width=.87\linewidth]{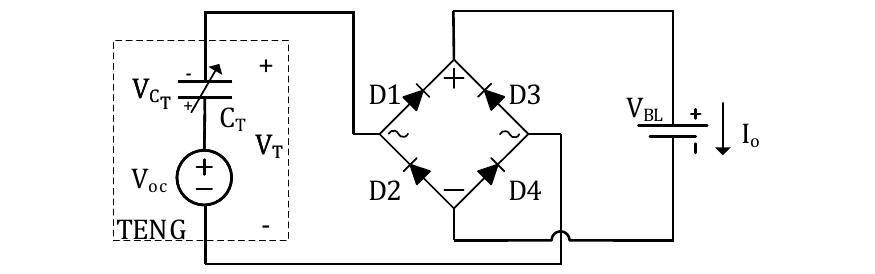}
  \vspace{-0.1in}
  \caption{Full Wave Rectifier (FWR) Circuit}\label{FWR}
  \end{center}
  \vspace*{-.25in}
\end{figure}

% ==== FIG 
% === V
The system starts at State~I, and initially, the TENG capacitor has zero charge and the TENG open-circuit voltage is also zero. As the two plates separate, the open-circuit voltage increases (linearly with separation, see (\ref{Voc})). The battery charging commences at the time when the TENG open-circuit voltage exceeds the battery voltage by two diode drops ($V_{oc}(t)\geq V_{BL}+2V_D$), and from that point on, the TENG voltage remains clamped to that same value, while the TENG capacitor accumulates charge equivalent to the difference $V_{oc}(t)-(V_{BL}+2V_D)$ until State II is reached. This completes the first half cycle. As the second half cycle begins, $V_{oc}$ starts to drop, causing the overall TENG voltage to fall below the threshold $V_{BL}+2V_D$ and the conduction stops. Battery no longer charges, and capacitor retains its charge to the level at State~II, until the time when a dual situation arises, namely, the absolute TENG voltage reaches $\abs{V_T(t)}\geq V_{BL}+2V_D$ and gets clamped at that value till State~I is reached, thus completing one full cycle (Refer Fig.~\ref{FWR_plot}).
\subsubsection{Per-Cycle Energy Output}The energy delivered to the load battery in each cycle denoted $E_{cycle}$, is the product of the battery voltage ($V_{BL}$) and the total charge flowing through it in that cycle ($\Delta Q_{cycle}$). From the charge conservation, this total charge flowing through the battery equals the change in the TENG capacitor charge in the two half-cycles, $\Delta Q_{cycle}=|\Delta Q_{C_T}^{I\rightarrow II}|+|\Delta Q_{C_T}^{II\rightarrow I}|$.   

By using (\ref{VTENG}), the voltage and the charge on the TENG capacitor at the two states, denoted $V_T^I,Q_{C_T}^I,V_T^{II},Q_{C_T}^{II}$, can be written as follows:
\vspace*{-.05in}
\begin{eqnarray}
&V^{I}_T=-\frac{Q_{C_T}^{I}}{C_{T,max}}=-(V_{BL}+2V_{D})\nonumber\\ 
&\Rightarrow Q_{C_T}^{I}=C_{T,max}(V_{BL}+2V_D) \label{eq:F1}\\
\vspace{-.05in}
&V^{II}_T=V_{oc,max}-\frac{Q_{C_T}^{II}}{C_{T,min}}=(V_{BL}+2V_{D})\nonumber\\
\vspace{-.05in}
&\Rightarrow Q_{C_T}^{II}=C_{T,min}(V_{oc,max}-V_{BL}-2V_D) \label{eq:F2}
\vspace*{-.05in}
\end{eqnarray}
It follows that,
\vspace*{-.05in}
\begin{equation}
\begin{split}
&\abs{\Delta Q^{I\rightarrow II}} = |Q_{C_T}^I-Q_{C_T}^{II}| \\
\vspace{-.05in}
=\mbox{ }& C_{T,min} V_{oc,max} -(C_{T,min}+C_{T,max})(V_{BL}+2V_{D})\\
\vspace{-.05in}
=\mbox{ }&\abs{\Delta Q^{II\rightarrow I}}=\abs{Q_{C_T}^{II}-Q_{C_T}^{I}}.\nonumber
\end{split}
\vspace*{-.05in}
\end{equation}
Since this same change in charge flows through the load battery in the respective half-cycles, the energy that the battery gains during one cycle is given by twice this change in charge times the battery voltage:
\vspace*{-.05in}
\begin{equation}
\label{eq:F4}
E_{cycle}  =2V_{BL}C_{T,min}[V_{oc,max}-(1+\beta)(V_{BL}+2V_{D})].
\vspace*{-.05in}
\end{equation}

\subsubsection{Optimal Battery Load}
From (\ref{eq:F4}), $E_{cycle}$ is a quadratic function of $V_{BL}$, and it can be maximized with respect to $V_{BL}$ by differentiating (\ref{eq:F4}) and setting it to zero, to yield the optimal battery load $V_{BL}^*$ and the optimal per-cycle energy, $E_{cycle}^*$:
\vspace*{-.05in}
\begin{eqnarray*}
\label{eq:F5}
V_{BL}^*&=&\frac{V_{oc,max}}{2(1+\beta)}-V_D\\
\label{eq:F6}
\vspace{-.05in}
E_{cycle}^* & =& \frac{C_{T,min}V_{oc,max}^2}{2(1+\beta)}\\
\vspace{-.05in}
&-&2C_{T,min}V_D(V_{oc,max}-(1+\beta)V_D).
\vspace*{-.05in}
\end{eqnarray*}
\subsubsection{Upper Bound to Battery Voltage}\label{FWR_UB}
In order for the load battery to charge, certain conditions must hold. In particular, for the charging in the first half cycle to commence, the maximum achievable voltage must exceed the upper threshold voltage needed for charging:
\vspace*{-.05in}
\begin{equation}
V_{oc,max}\geq V_{BL}+2V_D.\nonumber
\vspace*{-.05in}
\end{equation}
Similarly, for the charging in the second half cycle to occur, the minimum achievable voltage must fall below the lower threshold voltage required for charging:
\vspace*{-.05in}
\begin{equation}
-V_{C_T}^I\leq -(V_{BL}+2V_D).\nonumber
\vspace*{-.05in}
\end{equation}
Combining the two cases, it follows that for charging in both half-cycles, the following should hold:
\vspace*{-.05in}
\begin{equation}\label{eq:min}
\min\{V_{oc,max},V_{C_T}^I\}\geq V_{BL}+2V_D.
\vspace*{-.05in}
\end{equation}
Once the first half-cycle completes, the capacitor acquires a voltage of $V_{oc,max}-(V_{BL}+2V_D)$, and in the extreme case (corresponding to the upper bound battery load, where the conduction will occur just at the end of the second half-cycle, i.e., at State I), the conduction ceases until State~I, preserving the capacitor charge at: 
\vspace*{-.05in}
\[Q_{C_T}^{I}=Q_{C_T}^{II}=C_{T,min}[V_{oc,max}-(V_{BL}+2V_D)].\vspace*{-.05in}\]
So, 
\vspace*{-.15in}
\begin{eqnarray}
V_{C_T}^I&=&\frac{Q_{C_T}^{I}}{C_{T,max}}=\frac{Q_{C_T}^{II}}{C_{T,max}}\nonumber\\
\vspace{-.05in}
&=&\frac{C_{T,min}[V_{oc,max}-(V_{BL}+2V_D)]}{C_{T,max}}\nonumber\\
\vspace{-.05in}
&=&\frac{[V_{oc,max}-(V_{BL}+2V_D)]}{\beta}<V_{oc,max}.
\vspace*{-.05in}
\end{eqnarray}
Now since $V_{C_T}^I<V_{oc,max}$, (\ref{eq:min}) simplifies to:
\vspace*{-.05in}
\begin{eqnarray}
&&V_{C_T}^I\geq V_{BL}+2V_D\nonumber\\
\vspace{-.05in}
&\Leftrightarrow&\frac{[V_{oc,max}-(V_{BL}+2V_D)]}{\beta}\geq V_{BL}+2V_D\nonumber\\
\vspace{-.05in}
&\Leftrightarrow& V_{BL} \leq \frac{V_{oc,max}}{\beta+1}-2V_D.\label{Eq:15}
\vspace*{-.05in}
\end{eqnarray}
The last condition provides an upper bound to the load battery voltage for it to charge in the FWR extraction circuit. As expected of a quadratic function, it is double the optimal battery value.
% == 
\vspace*{-.15in}
\subsection{Parallel Synchronous Switched Harvesting on Inductor}
\begin{figure}[htbp]
\vspace*{-.1in}
  \begin{center}
  \includegraphics[width=.87\linewidth]{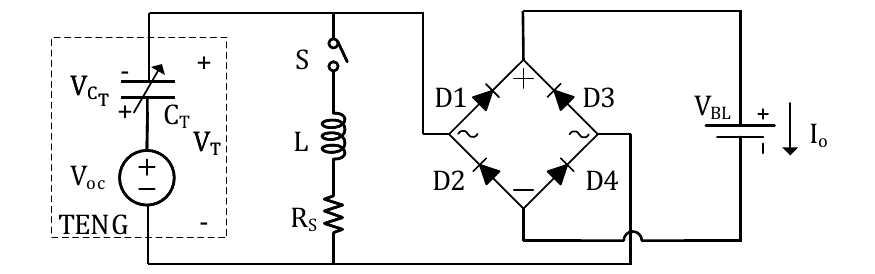}
  \vspace{-0.075in}
  \caption{Parallel Synchronous Switched Harvesting on Inductor (P-SSHI) Circuit}\label{PSSHI}
  \end{center}
  \vspace*{-.2in}
\end{figure}

\begin{figure}[htbp]
\vspace*{-.10in}
  \begin{center}
  \includegraphics[width=.87\linewidth]{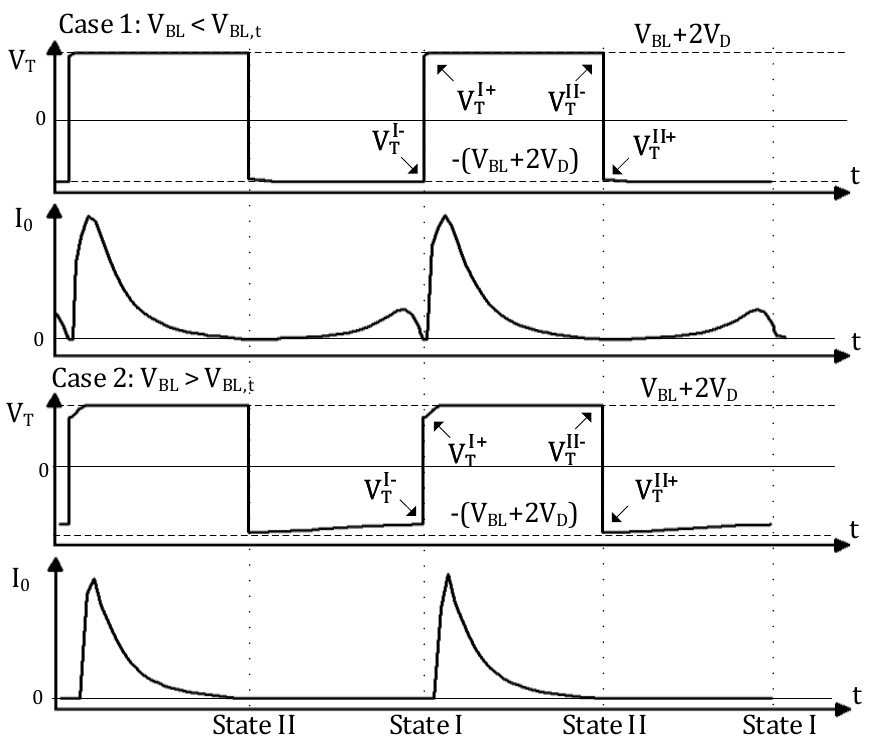}
  \vspace{-0.075in}
  \caption{Typical P-SSHI circuit TENG voltage: $V_T(t)$ and Output current: $I_o(t)$ for higher and lower load battery range.}\label{PSSHI_plot}
  \end{center}
  \vspace*{-.3in}
\end{figure}

\vspace*{-.05in}
P-SSHI circuit is shown in Fig.~\ref{PSSHI}. Compared to FWR, it has an added parallel path with a switch and an inductor.
The parallel switch S is closed at States I and II for half the $LC_T$ resonator time period, which flips TENG voltage polarity as apparent from the waveform plot in Fig.~\ref{PSSHI_plot}. 
This instant inversion increases the load charging duration (compared to the FWR circuit) in the following half-cycle. The voltage inversion is imperfect owing to the limited quality factor of the resonator $LC_T$ circuit being limited by the presence of switch on-state resistance ($R_{on}$) and the inductor parasitic resistance ($R_L$). Together those are modeled as $R_S$ in the following derivation of the per-cycle energy output ($E_{cycle}$). During the switch off states, the charging occurs similar to the FWR circuit with one difference. The TENG capacitor charge partly leaks through the parallel path of $S-L-R_s$ due to the finite off-state resistance ($R_{off}$) of the switch. Since the TENG voltage polarity and value, which dictate the extent of charge leakage, in the first half-cycle (separation of plates) and the second half-cycle (retraction of plates) are asymmetric, the corresponding leakage terms are also different, denoted $Q_L^I$ and $Q_L^{I}$ respectively in the following derivation.
\subsubsection{P-SSHI circuit analysis at States I and II}\label{VA}
For the sake of analytical convenience, we introduce four sets of notation for the pre and post switching TENG voltage ($V_T(t)$), TENG capacitor charge ($Q_{C_T}(t)$) and its voltage ($V_{C_T}(t)$) at States I and II:\\
$\bullet\!$ State$\!$ I- $ (V^{I-}_T\!,Q_{C_T}^{I-},V_{C_T}^{I-})\!\!\xrightarrow[]{T_{on}^{I}} \!\!$ State$\!$ I+ $(V_T^{I+}\!,Q_{C_T}^{I+},V_{C_T}^{I+})$\\
$\bullet\!$ State$\!$ II- $(V_T^{II-}\!\!,Q_{C_T}^{II-}\!\!,V_{C_T}^{II-})\!\!\xrightarrow[]{T_{on}^{II}} \!\!$ State$\!$ II+$(V_T^{II+}\!\!,Q_{C_T}^{II+}\!\!,V_{C_T}^{II+})$.
%\end{itemize} 

Here, $T_{on}^{I}$ and $T_{on}^{II}$ denote the switching periods at States I and II respectively, and equal the corresponding half the $LC_T$ resonator cycle time period. Those are designed to be small compared to the vibration cycle period ($T$); hence we can approximate the TENG capacitance $C_T(t)$ and open circuit voltage $V_{oc}(t)$ to be constant over the switching period.

\begin{figure}[htbp]
\vspace*{-.15in}
  \begin{center}
  \includegraphics[width=.87\linewidth]{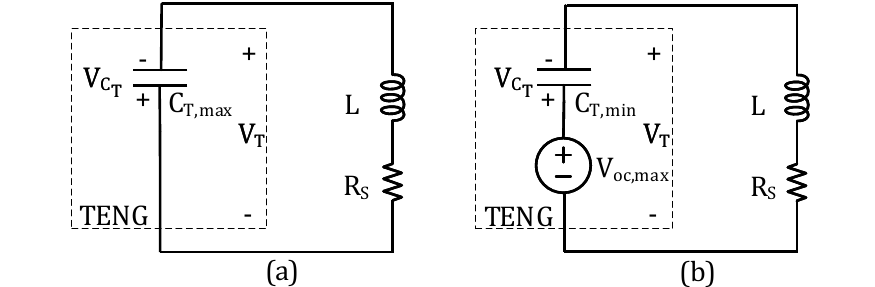}
  \vspace{-0.075in}
  \caption{Simplified P-SSHI circuits at States (a) I and (b) II}\label{PSSHI_simple}
  \end{center}
  \vspace*{-.2in}
\end{figure}

At State II, the P-SSHI circuit can be simplified to that in Fig.~\ref{PSSHI_simple}(b). We analyze it first to obtain the relation between pre and post switching TENG voltages and charges. 
The differential equation  for this resonator loop obtained using KVL is given by:
\vspace{-.05in}
\begin{equation}
%\label{eq:D1}
\frac{d^2V_{C_T}(t)}{dt}+\frac{R_S}{L}\frac{dV_{C_T}(t)}{dt}+\frac{V_{C_T}(t)}{LC_{T,min}}-\frac{V_{oc,max}}{LC_{T,min}}=0. \nonumber
\vspace{-.05in}
\end{equation}
This can be solved for TENG capacitor voltage $V_{C_T}(t)$ with initial condition as $V_{C_T}(0)=V_{C_T}^{II-}$ to obtain:
\vspace{-.05in}
\begin{equation}
\label{eq:D2}
\begin{split}
V_{C_T}(t)=&e^{-\frac{R_St}{2L}}(V_{C_T}^{II-}-V_{oc,max})(cos(\omega^{II}_{d}t)\\
\vspace{.05in}
&+\frac{R_S}{2L\omega^{II}_{d}}sin(\omega^{II}_{d}t))+V_{oc,max}, \nonumber
\end{split}
\vspace{-.05in}
\end{equation}
where $\omega^{II}_{d}$ is the resonance frequency given by,
\vspace{-.05in}
\begin{equation}
\label{eq:D3}
\omega^{II}_{d}:=\sqrt[]{\frac{1}{LC_{T,min}}-\frac{R_{S}^2}{4L^2}}.
\vspace{-.05in}
\end{equation}
Switch is kept ``on" for half the resonant time period ($T_{on}^{II}=\frac{\pi}{\omega^{II}_{d}}$). Hence,
\vspace{-.05in}
\begin{equation}
\label{eq:P2}
V_{C_T}^{II+}=V_{C_T}(\frac{\pi}{\omega^{II}_{d}})=-\alpha^{II}V_{C_T}^{II-}+(1+\alpha^{II})V_{oc,max},
\vspace{-.05in}
\end{equation}
with
\vspace{-.1in}
\begin{equation}
\label{eq:D4}
\alpha^{II}:=e^{\frac{-\pi R_S}{2\omega^{II}_{d}L}}=e^\frac{-\pi}{2Q^{II}_{f}},\mbox{ and } Q^{II}_{f}:=\frac{\omega^{II}_{d} L}{R_{S}}.
\vspace{-.05in}
\end{equation}
Here, $Q^{II}_{f}$ is the series resonator quality factor at State II, and $0<\alpha^{II}<1$ is its ``normalized" form.

The TENG voltage at the end of switching period ($V_T^{II+}$) can then be obtained from (\ref{eq:P2}) and the TENG equation (\ref{VTENG}):
\vspace{-.05in}
\begin{equation}
    \label{eq:P3}
    \begin{split}
    V_T^{II+}=V_{oc,max}-V_{C_T}^{II+} & =-\alpha^{II}V_T^{II-}.
    \end{split}
\vspace{-.05in}
\end{equation}
Thus, the normalized quality factor, $\alpha^{II}$ captures the reduction in the voltage swing (flip) at State II due to the circuit's series parasitic resistance ($R_{S}$).
On the other hand, during the switching at State I, the P-SSHI circuit is simplified to as shown in Fig.~\ref{PSSHI_simple}(a). Following the same process as above, we obtain:
\vspace{-.05in}
\begin{eqnarray}
    &V_T^{I+}=-V_{C_T}^{I+}=-\alpha^{I}V_T^{I-},\mbox{ where} \label{eq:P6} \\
    \vspace{-.05in}
    &\alpha^{I}:=e^{\frac{-\pi R_S}{2\omega^{I}_{d}L}}=e^\frac{-\pi}{2Q^{I}_{f}}; Q^{I}_{f}:=\frac{\omega^{I}_{d} L}{R_{S}} \label{eq:D6};\\
    \vspace{-.05in}
    &\omega^{I}_{d}:=\sqrt[]{\frac{1}{LC_{T,max}}-\frac{R_{S}^2}{4L^2}}.\label{eq:D5}
    \vspace{-.05in}
\end{eqnarray}
Note that the resonance frequency (\ref{eq:D5}) and the normalized quality factor (\ref{eq:D6}) have changed from that in State II, since the TENG capacitor value has changed to $C_{T,min}$ from $C_{T,max}$.

\subsubsection{Per-Cycle Energy Output}
Equations (\ref{eq:P3}) and (\ref{eq:P6}) related the post-switching TENG voltages to their pre-switching values at the States II and I, respectively. Next we derive the values for the pre-switching TENG voltages at the two States, to complete the characterization. This also allows us to then characterize the per-cycle energy output. As noted in the waveform plot (Fig.~\ref{PSSHI_plot}), similar to the FWR circuit, the TENG voltage is clamped at $V_{BL}+2V_D$  during the charging phase in the first half-cycle until the time State II is reached, which is when a charge reversal due to the switch closure occurs (for $T_{on}^{II}$ duration). Thus,
\vspace{-.05in}
\begin{eqnarray}
    &V_T^{II-}=V_{oc,max}-\frac{Q_{C_T}^{II-}}{C_{T,min}}=V_{BL}+2V_{D} \nonumber \\
    \vspace{-.05in}
    \Rightarrow&Q_{C_T}^{II-}=C_{T,min}(V_{oc,max}-V_{BL}-2V_{D}). \label{eq:P1}
    \vspace{-.05in}
\end{eqnarray}
Post switching at State II, using (\ref{eq:P3}),
\vspace{-.05in}
\begin{eqnarray}
    &V_T^{II+}&=V_{oc,max}-\frac{Q_{C_T}^{II+}}{C_{T,min}}=-\alpha^{II}(V_{BL}+2V_{D}) \nonumber \\
    \vspace{-.05in}
    \Rightarrow &Q_{C_T}^{II+}&=C_{T,min}(V_{oc,max}+\alpha^{II}(V_{BL}+2V_{D})). \label{eq:P4}
    \vspace{-.05in}
\end{eqnarray}

In the second half-cycle, there are two possibilities, either the TENG voltage reaches the lower threshold voltage, $-(2V_D+V_{BL})$ for charging of load to occur (Fig.~\ref{PSSHI_plot} Case 1) or the battery load is high enough to resist any current flow upto subsequent State I (Fig.~\ref{PSSHI_plot} Case 2). It is necessary to differentiate the two cases in order to obtain energy output, $E_{cycle}$ which is different for the two cases. 
% ==== FIG 

The ``transition" load value, $V_{BL,t}$ separating these two cases can be derived from the limiting case of no conduction in the second half cycle and TENG Voltage at State I ($V_T^{I-}$) just reaching the lower threshold, $-(V_{BL,t}+2V_D)$, i.e.,
\vspace{-.05in}
\begin{equation}
    \label{eq:P4A}
    V_T^{I-}=-\frac{Q_{C_T}^{I-}}{C_{T,max}}=-(V_{BL,t}+2V_{D}).
    \vspace{-.05in}
\end{equation}
Under the condition of no current flow through load in the second-half cycle, considering leakage, $Q_{C_T}^{I-}=Q_{C_T}^{II+}-\abs{Q_L^{II}}$. Then, inserting the value of $Q_{C_T}^{II+}$ from (\ref{eq:P4}) into (\ref{eq:P4A}), and invoking the definition of $\beta$, we arrive at:
\vspace{-.05in}
\begin{equation}
    \label{eq:P4B}
    V_{BL,t}=\frac{V_{oc,max}}{\beta-\alpha^{II}}-2V_D-\frac{\abs{Q_L^{II}}}{(\beta-\alpha^{II})C_{T,min}}.
    \vspace{-.05in}
\end{equation}
Case 1: {$V_{BL}<V_{Bl,t}$}: Here, TENG conducts in the second half-cycle, and hence, the TENG voltage is clamped to $-(V_{BL}+2V_{D})$ until reaching State I. Hence,
\vspace{-.05in}
\begin{eqnarray}
    &V_T^{I-}&=-\frac{Q_{C_T}^{I-}}{C_{T,max}}=-(V_{BL}+2V_{D}) \nonumber \\
    \vspace{-.05in}
    \Rightarrow&Q_{C_T}^{I-}&=C_{T,max}(V_{BL}+2V_{D}). \label{eq:P5}
    \vspace{-.05in}
\end{eqnarray}
Post switching at State I, using (\ref{eq:P6}),
\vspace{-.05in}
\begin{eqnarray}
    &V_T^{I+}&=-\frac{Q_{C_T}^{I+}}{C_{T,max}}=\alpha^{I}(V_{BL}+2V_{D}) \nonumber \\
    \vspace{-.05in}
    \Rightarrow &Q_{C_T}^{I+}&=-\alpha^{I}C_{T,max}(V_{BL}+2V_{D}). \label{eq:P7}
    \vspace{-.05in}
\end{eqnarray}
Current flows through the load barring the switching periods at States I and II. As before, the per-cycle energy delivered to the load ($E_{cycle}$) is the battery load ($V_{BL}$) times the charge flowing through it ($\Delta Q_{cycle}$). The absolute difference between the TENG capacitor charge at State I after switching ($Q_{C_T}^{I+}$) and the TENG capacitor charge at State II before switching ($Q_{C_T}^{II-}$) obtained using (\ref{eq:P7}) and (\ref{eq:P1}) respectively less the switch leakage charge between the above two states ($Q_L^I$) gives the total charge flowing through the load in the first half cycle, 
\vspace{-.05in}
\begin{equation}
\begin{split}
&\abs{\Delta Q^{I+ \rightarrow II-}}:=\abs{Q_{C_T}^{II-}-Q_{C_T}^{I+}}-\abs{Q_L^I}\\
\vspace{-.05in}
=\mbox{ }& C_{T,min} (V_{oc,max}+(\alpha^{I}\beta-1)(V_{BL}+2V_{D}))-\abs{Q_L^I}. \nonumber
\end{split}
\vspace{-.05in}
\end{equation}
Similarly, using (\ref{eq:P5}) and (\ref{eq:P4}) and taking into consideration the switch leakage charge between the States II+ and I- ($Q_L^{II}$), the charge flowing through the load in the second half cycle is given by,
\vspace{-.05in}
\begin{equation}
\begin{split}
&\abs{\Delta Q^{II+ \rightarrow I-}}:=\abs{Q_{C_T}^{I-}-Q_{C_T}^{II+}}-\abs{Q_L^{II}}\\
\vspace{-.05in}
=\mbox{ }& C_{T,min} (V_{oc,max}+(\alpha^{II}-\beta)(V_{BL}+2V_{D}))-\abs{Q_L^{II}}. \nonumber
\end{split}
\vspace{-.05in}
\end{equation}
Hence, $\Delta Q_{cycle}:=\abs{\Delta Q^{I+ \rightarrow II-}}+\abs{\Delta Q^{II+ \rightarrow I-}}$ and for this case, $E_{cycle}^1=V_{BL}\Delta Q_{cycle}$, implies:
\vspace{-.05in}
\begin{eqnarray}
\label{eq:P9}
E_{cycle}^1=\hspace{-15pt}&&V_{BL}C_{T,min}[2V_{oc,max}-[(1-\alpha^{I})\beta\\
\hspace{-15pt}&&+(1-\alpha^{II})](V_{BL}+2V_{D})]-V_{BL}(\abs{Q_L^I}+\abs{Q_L^{II}}).\nonumber
\vspace{-.05in}
\end{eqnarray}
%-------------------------
Case 2: $V_{BL} \geq V_{BL,t}$: Here, no conduction occurs through load in the second half cycle, implying $Q_{C_T}^{I-}=Q_{C_T}^{II+}-\abs{Q_L^{II}}$. So,
\vspace{-.05in}
\begin{equation}
\begin{split}
    V_T^{I-}&=-\frac{Q_{C_T}^{I-}}{C_{T,max}} \nonumber\\
    \vspace{-.05in}
    &=-\frac{V_{oc,max}+\alpha^{II}(V_{BL}+2V_{D})}{\beta}+\frac{\abs{Q_L^{II}}}{C_{T,max}} \nonumber
\end{split}
\vspace{-.05in}
\end{equation}
Post switching at State I, using (\ref{eq:P6}),
\vspace{-.05in}
\begin{eqnarray}
    V_T^{I+}&=&-\frac{Q_{C_T}^{I+}}{C_{T,max}}\nonumber\\
    \vspace{-.05in}
    &=&\frac{\alpha^{I}}{\beta}(V_{oc,max}+\alpha^{II}(V_{BL}+2V_{D}))-\frac{\alpha^I\abs{Q_L^{II}}}{C_{T,max}}\nonumber\\
    \vspace{-.05in}
    \Rightarrow Q_{C_T}^{I+}&=&-\alpha^{I}C_{T,min}(V_{oc,max}+\alpha^{II}(V_{BL}+2V_{D}))\nonumber\\
    \vspace{-.05in}
    &&+\alpha^{I}\abs{Q_L^{II}}.\label{eq:P11}
    \vspace{-.05in}
\end{eqnarray}
From (\ref{eq:P1}) and (\ref{eq:P11}) and considering the leakage charge, $Q_L^I$ we find the charge flowing through the load in the first half cycle is given by,
\vspace{-.05in}
\begin{equation}
\label{eq:P12}
\begin{split}
&\abs{\Delta Q^{I+ \rightarrow II-}}=\abs{Q_{C_T}^{II-}-Q_{C_T}^{I+}}-\abs{Q_L^I}\\
\vspace{-.05in}
=\mbox{ }& C_{T,min} [(1+\alpha^{I})V_{oc,max}-(1-\alpha^{I}\alpha^{II})(V_{BL}+2V_{D})]\nonumber\\
\vspace{-.05in}
&-\alpha^{I}\abs{Q_L^{II}}-\abs{Q_L^I}. \nonumber
\end{split}
\vspace{-.05in}
\end{equation}
Since in this case, $\Delta Q_{cycle}=\abs{\Delta Q^{I+ \rightarrow II-}}$, we can obtain,
\vspace{-.05in}
\begin{eqnarray}
E_{cycle}^2&=&V_{BL}C_{T,min}[(1+\alpha^{I})V_{oc,max}\nonumber\\
&&-(1-\alpha^{I}\alpha^{II})(V_{BL}+2V_{D})]\nonumber\\
&&-V_{BL}(\abs{Q_L^I}+\alpha^I\abs{Q_L^{II}}).\label{eq:P13}
\end{eqnarray}
\subsubsection{Optimal Battery Load}
We first show that the optimal condition occurs in the case when, $V_{BL} \geq V_{BL,t}$. Using (\ref{eq:P4B}):
\vspace{-.05in}
\begin{eqnarray}\label{eq:P13B}
V_{BL}&\geq&\hspace{-6pt}V_{BL,t}=\frac{V_{oc,max}}{\beta-\alpha^{II}}-2V_D-\frac{\abs{Q_L^{II}}}{(\beta-\alpha^{II})C_{T,min}}\nonumber\\
\vspace{-.05in}
&\Leftrightarrow&\hspace{-6pt}(\beta-\alpha^{II})(V_{BL}+2V_D)-V_{oc,max}+\frac{\abs{Q_L^{II}}}{C_{T,min}}\geq0.\nonumber
\vspace{-.05in}
\end{eqnarray}
Also from (\ref{eq:P13}) and (\ref{eq:P9}),
\vspace{-.05in}
\begin{eqnarray}
    &&E_{cycle}^{2}-E_{cycle}^{1}=V_{BL}(1-\alpha^{I})\nonumber\\
    \vspace{-.05in}
    &&[C_{T,min}((\beta-\alpha^{II})(V_{BL}+2V_D)-V_{oc,max})+\abs{Q_L^{II}}]\nonumber
    \vspace{-.05in}
\end{eqnarray}
Employing the above inequality into the last equation, and noting that $V_{BL}>0$ and $0<\alpha^{II}<1$, it follows that $E_{cycle}^2-E_{cycle}^1\geq 0$, implying that a larger per-cycle energy is harvested when $V_{BL} \geq V_{BL,t}$. The optimal battery voltage ($V_{BL}^*$) and the maximized per-cycle energy ($E_{cycle}^*$) can then be found by analyzing this case to yield:
\vspace{-.05in}
\begin{eqnarray}
V_{BL}^*&=&\hspace{-4pt}\frac{1}{2(1-\alpha^{I}\alpha^{II})}[(1+\alpha^{I})V_{oc,max}-\frac{c}{C_{T,min}}]-V_{D} \nonumber\\ 
\vspace{-.05in}
E_{cycle}^*&=&\hspace{-4pt}\frac{(1+\alpha^{I})^2}{4(1-\alpha^{I}\alpha^{II})}(C_{T,min}V_{oc,max}^2)\nonumber\\
\vspace{-.05in}
&&\hspace{-4pt}-C_{T,min}V_D[(1+\alpha^{I})V_{oc,max}-(1-\alpha^{I}\alpha^{II})V_D]\nonumber\\
\vspace{-.05in}
&&\hspace{-4pt}+\frac{c}{2(1-\alpha^{I}\alpha^{II})}\left[\frac{c}{2C_{T,min}}-(1+\alpha^{I})V_{oc,max}\right]\nonumber\\
\vspace{-.05in}
&&\hspace{-4pt}+cV_D; \hspace{5pt}c=\abs{Q_L^I}+\alpha^I\abs{Q_L^{II}}.\nonumber
\vspace{-.05in}
\end{eqnarray}
As detailed in Sec.\ref{Results}, the values of the non-ideality parameters ($V_D$, $\alpha^{I}$/$\alpha^{II}$, and $Q_{L}^{I}$/$Q_{L}^{II}$) change significantly over the operation range as the load battery varies, causing the currents and voltages in the circuits to change. Hence, to estimate the optimal battery load and corresponding optimal energy output, their averaged values over the entire load operating range are used in the above derived equations. 
%%%%
\subsubsection{Upper bound to Battery Voltage}
If the load $V_{BL}$ is higher than $V_{oc,max}-2V_D-\frac{Q_L^I}{C_{T,min}}$, there shall be transient operation cycles before the steady-state charging commences. We introduce the notation $V_{T,k}$ to denote the TENG voltage in the $k$th cycle. Then starting from rest, i.e., $V_{T,1}^{I}=0$ and no conduction through the load in the first half-cycle, the TENG voltage at State II is simply given by,
\vspace{-.05in}
\begin{eqnarray*}
V_{T,1}^{II-}=V_{oc,max}-V_L^I;\hspace{2pt}V_{L}^{I}=\frac{\abs{Q_L^I}}{C_{T,min}}.
\vspace{-.05in}
\end{eqnarray*}
which is then followed by a voltage inversion due to the switching action, implying,
\vspace{-.05in}
\begin{eqnarray*}
V_{T,1}^{II+}=-\alpha^{II}(V_{oc,max}-V_L^I).
\vspace{-.05in}
\end{eqnarray*}
Similarly, when no conduction occurs in the second half-cycle,
\vspace{-.05in}
\begin{eqnarray*}
V_{T,2}^{I-}=-\frac{(1+\alpha^{II})}{\beta}V_{oc,max}+\frac{\alpha^{II}}{\beta}V_L^I+V_L^{II};\hspace{2pt} V_{L}^{II}=\frac{\abs{Q_L^{II}}}{C_{T,max}}.
\vspace{-.05in}
\end{eqnarray*}
The voltage is flipped again at the onset of the second vibration cycle, implying,
\vspace{-.05in}
\begin{eqnarray*}
\hspace{-10pt}&\Rightarrow&\hspace{-10pt} V_{T,2}^{I+}=\frac{\alpha^{I}(1+\alpha^{II})}{\beta}V_{oc,max}-\frac{\alpha^{I}\alpha^{II}}{\beta}V_L^I-\alpha^{I}V_L^{II};\\
\hspace{-10pt}&\Rightarrow&\hspace{-10pt} V_{T,2}^{II-}=(1+\alpha^{I}+\alpha^{I}\alpha^{II})V_{oc,max}-(1+\alpha^{I}\alpha^{II})V_L^I\\
\hspace{-10pt}&&\hspace{24pt}-\alpha^{I}\beta V_L^{II}; \\
\hspace{-10pt}&\Rightarrow&\hspace{-10pt} V_{T,3}^{II-}=\left[ \{ 1+\alpha^{I}\alpha^{II}+(\alpha^{I}\alpha^{II})^2 \}+\alpha^{I} \{ 1+\alpha^{I}\alpha^{II} \} \right]\\
\hspace{-10pt}&&\hspace{24pt}V_{oc,max}-\{ 1+\alpha^{I}\alpha^{II}+(\alpha^{I}\alpha^{II})^2 \}V_L^{I}\\
\hspace{-10pt}&&\hspace{24pt}-\alpha^{I}(1+\alpha^{I}\alpha^{II})\beta V_L^{II}. 
\end{eqnarray*}
Following the above reasoning, the maximum TENG voltage in the $k^{th}$ cycle (at State II) can be written as a sum of geometric series with common ratio $\alpha^{I}\alpha^{II}$, i.e., 
\vspace{-.05in}
\begin{eqnarray*}
\hspace{-10pt}&&\hspace{-10pt}V_{T,k}^{II-}=\left[ \frac{1-(\alpha^{I}\alpha^{II})^k}{1-\alpha^{I}\alpha^{II}}+\alpha^{I}\frac{1-(\alpha^{I}\alpha^{II})^{k-1}}{1-\alpha^{I}\alpha^{II}} \right] V_{oc,max}\\
\hspace{-10pt}&&\hspace{24pt}-\frac{1-(\alpha^{I}\alpha^{II})^{k}}{1-\alpha^{I}\alpha^{II}}  V_{L}^I-\frac{1-(\alpha^{I}\alpha^{II})^{k-1}}{1-\alpha^{I}\alpha^{II}}(\alpha^I\beta V_L^{II}).
\end{eqnarray*}
Thus, through the switching at the extremes, the TENG voltage will continue to build up until it reaches $V_{BL}+2V_D$, post to which charging commences and the energy output equation ($E_{cycle}$) derived in (\ref{eq:P13}) yields the per-cycle harvested energy. 

An upper bound to the battery voltage can be derived from the extreme case where no charging occurs even as $k\rightarrow\infty$, and with $\alpha^{I}\alpha^{II}<1$, we get,
\vspace{-.05in}
\begin{eqnarray*}
    V_{BL}\leq \lim_{k\rightarrow\infty} V_{T,k}^{II-}-2V_D \hspace{-10pt}&=&\hspace{-10pt}\frac{(1+\alpha^{I})V_{oc,max}-V_L^I-\alpha^I\beta V_L^{II}}{1-\alpha^{I}\alpha^{II}}\\
    \hspace{-10pt}&&\hspace{-10pt}-2V_D.
    \vspace{-.05in}
\end{eqnarray*}
%--------------------------------------------------------------------------
\vspace*{-.35in}
\subsection{Series Synchronous Switched Harvesting on Inductor}
\begin{figure}[htbp]
\vspace*{-.1in}
  \begin{center}
  \includegraphics[width=1\linewidth]{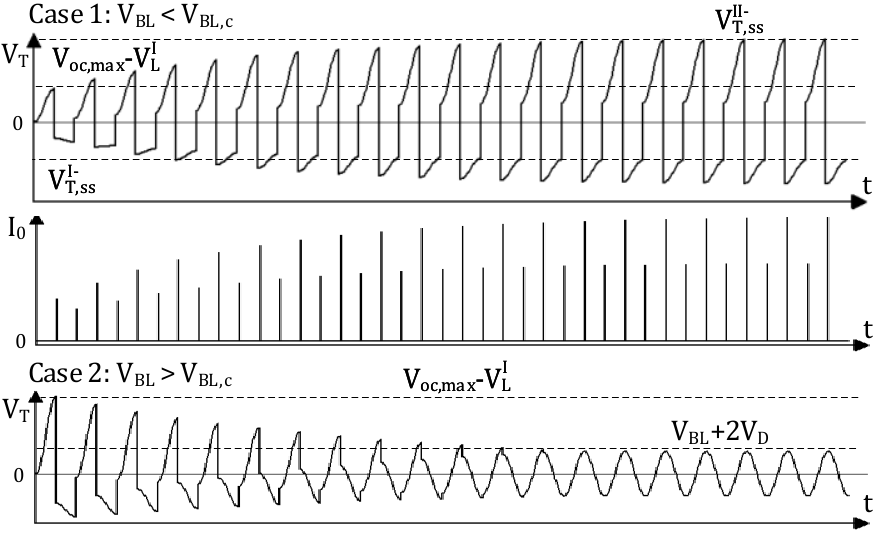}
  \vspace{-0.25in}
  \caption{(Top) Typical S-SSHI circuit TENG voltage: $V_T(t)$ and Output current: $I_o(t)$ for below critical battery load and (Bottom) TENG voltage: $V_T(t)$ for above critical battery load.}\label{SSSHI_plot}
  \end{center}
 \vspace*{-.2in}
\end{figure}

\vspace*{-.05in}
While TENGs generally have high open circuit voltage ($V_{oc,max}$), in the earlier discussed circuits, the maximum TENG voltage ($V_T$) gets clamped to a value limited by the battery load voltage. In contrast, a serial switch can allow the voltage buildup irrespective of the battery load, as in the presented case of the S-SSHI circuit.
\begin{figure}[htbp]
  \vspace*{-.1in}
  \begin{center}
  \includegraphics[width=.87\linewidth]{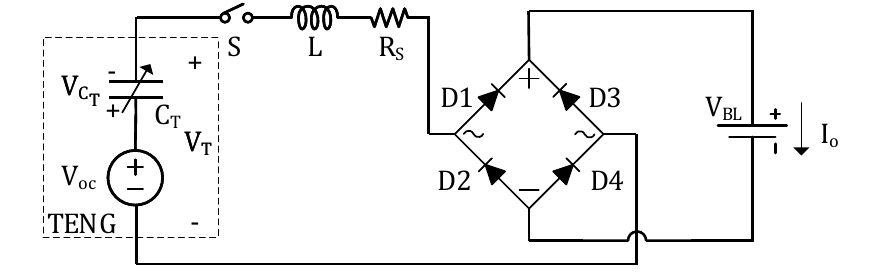}
  \vspace{-0.075in}
  \caption{Series Synchronous Switched Harvesting on Inductor (S-SSHI) Circuit}\label{S-SSHI}
  \end{center}
  \vspace*{-.2in}
\end{figure}
It consists of a serial switch connecting TENG with the battery load via an inductor $L$ (Fig.~\ref{S-SSHI}). The TENG operates in an open circuit condition (barring the leakage current) except when the switch is closed at States I and II for half the $LC_T$ resonator cycle to flip the TENG voltage polarity and in that process charges the battery (Fig.~\ref{SSSHI_plot}). For the battery to be charged, the battery voltage must be below a critical value (derived below in Section~\ref{S-SSHI_upper}). Fig.~\ref{SSSHI_plot} contrasts the two cases, one where this condition is satisfied (top two plots, where the TENG voltage builds up and the battery charging current are shown) versus where the condition is not satisfied (bottom plot, where the TENG voltage does not build up and no battery charging current flows).
\subsubsection{S-SSHI circuit analysis at States I II}
\begin{figure} [htbp]
  \vspace*{-.05in}
  \begin{center}
  \includegraphics[width=.87\linewidth]{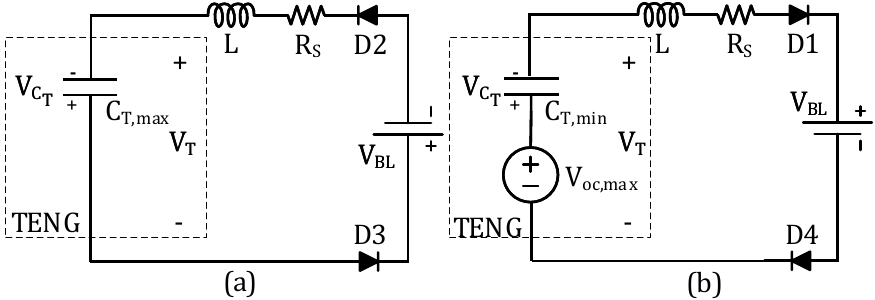}
  \vspace*{-0.075in}
  \caption{Simplified S-SSHI circuits at States (a) I and (b) II}\label{SSSHI_simple}
  \end{center}
  \vspace*{-.25in}
\end{figure}

We first analyze the simplified S-SSHI circuit at State II, as in Fig.~\ref{SSSHI_simple}(b), to obtain the relation between and pre and post switched TENG voltages.
The initial voltage of the TENG capacitor before the switching at Stage II is still denoted as $V_{C_T}^{II-}$. Using KVL around the loop, we arrive at the differential equation below,
\vspace{-.05in}
\begin{equation}
\label{eq:A1}
\frac{d^2V_{C_T}(t)}{dt}+\frac{R_S}{L}\frac{dV_{C_T}(t)}{dt}+\frac{V_{C_T}(t)}{LC_{T,min}}=\frac{V_{oc,max}-V_{BL}-2V_D}{LC_{T,min}}.
\vspace{-.05in}
\end{equation}
As in the case of the P-SSHI circuit, the switch is closed for half the $LC_T$ resonator cycle. Accordingly, solving the above differential equation and performing the analysis as in Section \ref{VA}, the TENG capacitor voltage after the switching ($V_{C_T}^{II+}$) is given by, 
\vspace{-.05in}
\begin{eqnarray}
&V_{C_T}^{II+}=-\alpha^{II}V_{C_T}^{II-}+(1+\alpha^{II})(V_{oc,max}-V_{BL}-2V_D).\nonumber
\vspace{-.05in}
\end{eqnarray}
Using the TENG equation (\ref{VTENG}),
\vspace{-.05in}
\begin{eqnarray}
V_{T}^{II+}=-\alpha^{II}V_{T}^{II-}+(1+\alpha^{II})(V_{BL}+2V_D).\label{eq:A3}
\vspace{-.05in}
\end{eqnarray}

On the other hand, at State I, the S-SSHI circuit is simplified to that in Fig.~\ref{SSSHI_simple}(a), with pre and post switching TENG voltages related by (obtained by solving an equation of the type (\ref{eq:A1}) with $V_{oc,max}$ replaced by $V_{oc,min}=0$ so that $V_{C_T}=-V_T$, and $C_{T,min}$ replaced by $C_{T,max}$),
\vspace{-.05in}
\begin{eqnarray}
V_{T}^{I+}=-\alpha^{I}V_{T}^{I-}-(1+\alpha^{I})(V_{BL}+2V_D).\label{eq:A5}
\vspace{-.05in}
\end{eqnarray}
Here, $\alpha^{II}$ and $\alpha^{I}$ are the same as those defined earlier in (\ref{eq:D4}) and (\ref{eq:D6}), respectively. 
\subsubsection{Switching among States: Transient Phase}
The previous subsection discussed the S-SSHI circuit when the switch is closed at States I and II and derived the relations among the pre-switching and post-switching TENG voltages at both those states. Next we derive the pre-switching TENG voltages at the two states to complete the characterization of the TENG voltages. Unlike P-SSHI, the S-SSHI circuit has a transient phase and either the TENG voltage settles to a steady-state value higher than or equal to $V_{oc,max}$, or a value less than $V_{BL}+2V_D$  (refer Fig.~\ref{SSSHI_plot}), depending on whether the battery voltage is below or above a critical value. We model the TENG voltage transients to derive the steady-state values at States I and II. As before, we let the $V_{T,k}$ denote the TENG voltage in the $k^{th}$ cycle. 

With no current flow other than leakage between States I and II with switch S open, $Q_{C_T,k}^{II-}=Q_{C_T,k}^{I+}-\abs{Q_L^I}$. Hence,
\vspace{-.05in}
\begin{equation}
\label{eq:S2}
\begin{split}
     V_{T,k}^{II-}=V_{oc,max}-\frac{Q_{C_T,k}^{II-}}{C_{T,min}}=V_{oc,max}+\beta V_{T,k}^{I+}-V_L^I.
     \end{split}
     \vspace{-.05in}
\end{equation}
Using the above equation and the relation in (\ref{eq:A3}), the post-switching TENG voltage at State II is given by,
\vspace{-.05in}
\begin{equation}
\label{eq:S3}
\begin{split}
     V_{T,k}^{II+}=V_{oc,max}-\frac{Q_{C_T,k}^{II+}}{C_{T,min}}=&-\alpha^{II}(\beta V_{T,k}^{I+}+V_{oc,max}-V_L^I)\\
     \vspace{-.05in}
     &+(1+\alpha^{II})(V_{BL}+2V_D).\nonumber
     \end{split}
     \vspace{-.05in}
\end{equation}
With only leakage current up to reaching State I of the subsequent $(k+1)^{th}$ cycle ($Q_{T,k+1}^{I-}=Q_{T,k}^{II+}-\abs{Q_L^{II}}$) and from above equation, it follows, 
\vspace{-.05in}
\begin{equation}
    \label{eq:S4}
\begin{split}
     V_{T,k+1}^{I-}=-\frac{Q_{C_T,k+1}^{I-}}{C_{T,max}}&=-\alpha^{II}V_{T,k}^{I+}+\frac{\alpha^{II}V_L^I}{\beta}+V_L^{II}\\
     \vspace{-.05in}
     &-\frac{(1+\alpha^{II})}{\beta}(V_{oc,max}-V_{BL}-2V_D).\nonumber
     \end{split}
     \vspace{-.05in}
\end{equation}
Using (\ref{eq:A5}), we arrive at the desired relation,
\vspace{-.05in}
\begin{eqnarray}
    V_{T,k+1}^{I+}&=&c_1V_{T,k}^{I+}+c_2,\mbox{ where} \nonumber \\
    \vspace{-.05in}
    c_1&:=&\alpha^{I}\alpha^{II}; \nonumber \\
    \vspace{-.05in}
    c_2&:=&-[1+\alpha^{I}+\frac{\alpha^{I}(1+\alpha^{II})}{\beta}](V_{BL}+2V_D)\nonumber\\
    \vspace{-.05in}
    &+&\frac{\alpha^{I}(1+\alpha^{II})}{\beta}V_{oc,max}-\frac{\alpha^{I}\alpha^{II}V_L^I}{\beta}-\alpha^{I}V_L^{II}.\nonumber\\
    \label{eq:J0}
    \vspace{-.05in}
\end{eqnarray}
Above is a recursive relation, which can be solved to obtain the relation between $V_{T,k+1}^{I+}$ and $V_{T,1}^{I+}$:
\vspace{-.05in}
\begin{equation}
     V_{T,k}^{I+}= c_1^{k-1}V_{T,1}^{I+}+c_1^{k-2}c_2+....+c_2. \nonumber
     \vspace{-.05in}
\end{equation}
Taking the TENG operation to start from rest, i.e., $V_{T,1}^{I+}=0$, the above equates to a sum of geometric series with $(k-1)$ terms and common ratio, $c_1=\alpha^{I}\alpha^{II}<1$, and the first term as $c_2$. It follows that,
\vspace{-.05in}
\begin{equation}
    \label{eq:S5}
     V_{T,k}^{I+}= \frac{1-c_1^{k-1}}{1-c_1}c_2.
     \vspace{-.05in}
\end{equation}
This provides a closed-form solution for the post-switching TENG voltage at State I of the $k$th cycle. Also using (\ref{eq:S2}),
\vspace{-.05in}
\begin{equation}
    \label{eq:S6}
     V_{T,k}^{II-}= V_{oc,max}+\beta\frac{1-c_1^{k-1}}{1-c_1}c_2-V_L^I.
     \vspace{-.05in}
\end{equation}
From above equation it is clear that when $c_2>0$, the TENG voltage builds up over time (increases with respect to $k$) as in Fig.~\ref{SSSHI_plot} Case 1, and otherwise when $c_2<0$, it diminishes over time as in Fig.~\ref{SSSHI_plot} Case 2. The latter case leads to no charging in the steady-state. No transient cycles are present for the special case of $c_2=0$. 
\subsubsection{Per-Cycle Energy Output in Steady-State}
In the steady-state, from (\ref{eq:S5}), the post-switching voltage and charge on TENG at State I can simply be written as:
\vspace{-.05in}
\begin{eqnarray}
    V_{T,ss}^{I+}&=&\frac{c_2}{1-c_1}=-\frac{Q_{C_T,ss}^{I+}}{C_{T,max}} \nonumber \\
    \vspace{-.05in}
    \Rightarrow Q_{C_T,ss}^{I+}&=&-C_{T,max}\frac{c_2}{1-c_1}.\label{eq:J1}
    \vspace{-.05in}
\end{eqnarray}
Using this, along with the relations (\ref{eq:S2})-(\ref{eq:S4}), we obtain the TENG capacitor charge values in the steady-state as:
\vspace{-.05in}
\begin{eqnarray}
    Q_{C_T,ss}^{II-}\hspace{-2pt}&=&\hspace{-2pt}Q_{C_T,ss}^{I+}-\abs{Q_L^I};\label{eq:J2}\\
    \vspace{-.05in}
   Q_{C_T,ss}^{II+}\hspace{-2pt}&=&\hspace{-2pt}C_{T,min}[\frac{\alpha^{II}\beta c_2}{1-c_1}\label{eq:J3}\\
   \vspace{-.05in}
    \hspace{-2pt}&+&\hspace{-2pt}(1+\alpha^{II})(V_{oc,max}-V_{BL}-2V_D)]-\alpha^{II}\abs{Q_L^I};\nonumber\\
    \vspace{-.05in}
   Q_{C_T,ss}^{I-}\hspace{-2pt}&=&\hspace{-2pt}Q_{C_T,ss}^{II+}-\abs{Q_L^{II}}.\label{eq:J4}
   \vspace{-.05in}
\end{eqnarray}
The change in TENG capacitor charge during the two switching periods flows through the load. S-SSHI being a series circuit, the leakage charge flows through the load too. So the charge flowing through the load in one cycle equals,
\vspace{-.05in}
\begin{equation}
\label{eq:S16}
\begin{split}
\Delta Q_{cycle}\hspace{-3pt}&=\abs{\Delta Q^{I-\rightarrow I+}}+\abs{\Delta Q^{II-\rightarrow II+}}+\abs{Q_L}\nonumber\\
\vspace{-.05in}
\hspace{-3pt}&=\abs{Q_{T,ss}^{I+}-Q_{T,ss}^{I-}}+\abs{Q_{T,ss}^{II-}-Q_{T,ss}^{II+}}+\abs{Q_L^I}+\abs{Q_L^{II}}.\nonumber
\end{split}
\vspace{-.05in}
\end{equation}
Hence, the per-cycle energy output equals, $V_{BL}\Delta Q_{cycle}$. Using (\ref{eq:J1}-\ref{eq:J4}) and inserting value of $c_1$ and $c_2$ from (\ref{eq:J0}),
\vspace{-.05in}
\begin{eqnarray}
\label{eq:S17}
&&E_{cycle}=\left[\frac{2(1+\alpha^{I})(1+\alpha^{II})}{(1-\alpha^{I}\alpha^{II})}\right]C_{T,min}V_{BL}\\
\vspace{-.05in}
&&\times\left[V_{oc,max}-(1+\beta)(V_{BL}+2V_D)\right]-\frac{2V_{BL}}{(1-\alpha^{I}\alpha^{II})}c_3;\nonumber\\
\vspace{-.05in}
&&c_3=(\alpha^{II}(1+2\alpha^{I})-1)\abs{Q_L^I}+\alpha^{I}(1+\alpha^{II})\abs{Q_L^{II}}.\nonumber
\vspace{-.05in}
\end{eqnarray}
The number of transient cycles required to attain the steady-state depends on the product of the two normalized quality factors $c_1=\alpha^{I}\alpha^{II}$ (or equivalently their geometric mean). A system with a lower quality factor shall reach the steady-state quicker but with a lower per-cycle energy output for a given load.
\subsubsection{Optimal Battery Load}
$E_{cycle}$ found in (\ref{eq:S17}) is quadratic in $V_{BL}$, which can be optimized to attain the optimal battery load ($V_{BL}^*$) and maximized per-cycle energy ($E_{cycle}^*$):
\vspace{-.05in}
\begin{eqnarray*}
V_{BL}^*&=&\frac{V_{oc,max}}{2(1+\beta)}-V_D\\
\vspace{-.05in}
&&-\frac{c_3}{2C_{T,min}(1+\beta)(1+\alpha^{I})(1+\alpha^{II})}\\
\vspace{-.05in}
E_{cycle}^*&=&\left[\frac{2C_{T,min}(1+\alpha^{I})(1+\alpha^{II})}{(1-\alpha^{I}\alpha^{II})}\right]\\
\vspace{-.05in}
&\times&\left[\frac{V_{oc,max}^2}{4(1+\beta)}-V_{oc,max}V_D+(1+\beta)V_D^2\right]\\
\vspace{-.05in}
&-&\frac{c_3}{(1-\alpha^{I}\alpha^{II})}\left[\frac{V_{oc,max}}{\beta+1}-2V_D\right.\\
\vspace{-.05in}
&&\hspace{1pt}\left.-\frac{c_3}{2C_{T,min}(1+\beta)(1+\alpha^{I})(1+\alpha^{II})}\right]
\vspace{-.05in}
\end{eqnarray*}
\subsubsection{Upper bound to Battery Voltage}\label{S-SSHI_upper}
As discussed earlier, condition for S-SSHI circuit to reach a non-zero steady-state (so battery can be charged in steady-state) is:
\vspace{-.05in}
\begin{eqnarray}
&&c_2\geq 0 \nonumber\\
\vspace{-.05in}
&\Leftrightarrow&V_{BL}\leq \frac{V_{oc,max}}{1+\frac{(1+\alpha^{I})\beta}{\alpha^{I}(1+\alpha^{II})}}-\frac{\alpha^{I}\alpha^{II}V_L^I}{(1+\alpha^{I})(1+\beta)}\nonumber\\
\vspace{-.05in}
&&\hspace*{.25in}-\frac{V_{L}^{II}}{1+\frac{1}{\alpha^I}+\frac{(1+\alpha^{II})}{\beta}}-2V_D.\label{eq:S18}
\vspace{-.05in}
\end{eqnarray}
The above condition sets an upper bound to the battery voltage for a sustained S-SSHI charging.
\vspace*{-.15in}
\subsection{Comparative Analysis}
\vspace*{-.05in}
In the above sections, we derived the closed-form results for each of the three considered circuits' per-cycle energy output. Here, we compare their optimal output in a common analytical framework. For the sake of simplicity, we relax the analysis conditions by using ideal diodes and switches, i.e., take the values of diode voltage drop ($V_D$), leakage ($Q_L^I/Q_L^{II}$), and switch-on resistance ($R_{on}$) as zero. For further convenience, we define:
\vspace*{-.05in}
\begin{equation}
E_{ref}=C_{T,min}V_{oc,max}^{2}.\nonumber
\vspace*{-.05in}
\end{equation} 
This allows easy tabulation of optimal per-cycle energy output ($E^*_{cycle}$), optimal battery load ($V^*_{BL}$), and maximum gain over FWR circuit ($\frac{E^*_{cycle}}{E^*_{FWR,cycle}}$) as in Table~\ref{tab:optimal}.
\begin{table}[ht]
\vspace*{-.1in}
    \centering
    \caption{Optimal energy output, optimal battery load, and maximum gain over FWR (under $V_D\!=\!0$; $Q_L^{I}\!=\!Q_L^{II}\!=\!0$)}
    \label{tab:optimal}
    \begin{tabular}{c|c|c|c}
    \hline
    Circuit &$\frac{V^*_{BL}}{V_{oc,max}}$  &$\frac{E^*_{cycle}}{E_{ref}}$ &$\frac{E^*_{cycle}}{E^*_{FWR,cycle}}$ \\[2ex]
    \hline
    
    FWR &$\frac{1}{2(1+\beta)}$ &$\frac{1}{2(1+\beta)}$  &1  \\[2ex]
    
    P-SSHI &$\frac{(1+\alpha^{I})}{2(1-\alpha^{I}\alpha^{II})}$ &$\frac{(1+\alpha^{I})^{2}}{4(1-\alpha^{I}\alpha^{II})}$  &$\frac{(1+\alpha^{I})^{2}(1+\beta)}{2(1-\alpha^{I}\alpha^{II})}$  \\[2ex]
    
    S-SSHI  &$\frac{1}{2(1+\beta)}$ &$\frac{(1+\alpha^{I})(1+\alpha^{II})}{2(1-\alpha^{I}\alpha^{II})(1+\beta)}$ &$\frac{(1+\alpha^{I})(1+\alpha^{II})}{(1-\alpha^{I}\alpha^{II})}$  \\[2ex]
    \hline
    \end{tabular}
    \vspace*{-0.2in}
\end{table}

\begin{figure} [htbp]
  \vspace*{-.10in}
  \begin{center}
  \includegraphics[width=1\linewidth]{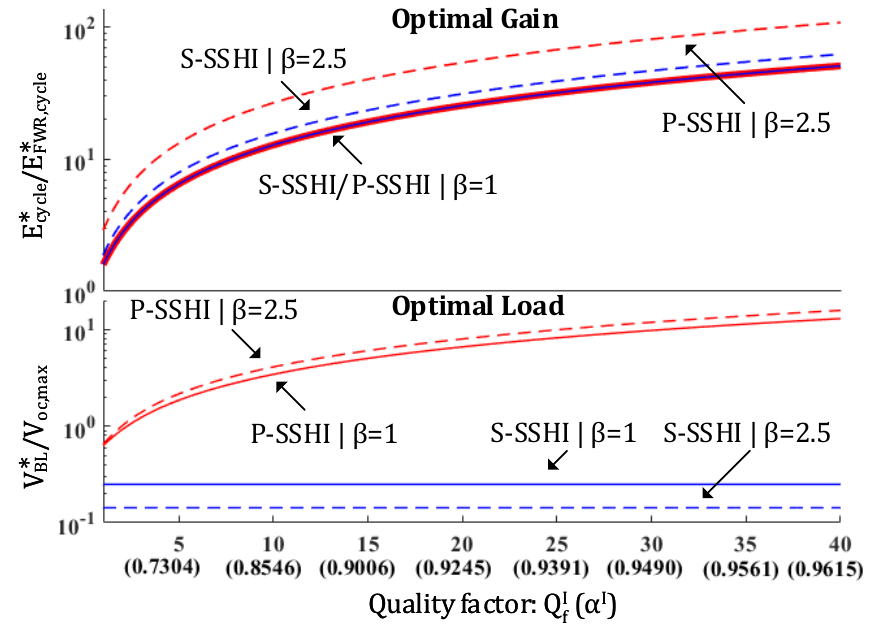}
  \vspace*{-.175in}
  \caption{Top/Bottom: Comparison of optimal energy gain (resp., battery load) of the P-SSHI and S-SSHI circuits over FWR (resp., over $V_{oc,max}$) against the inductor quality factor.}\label{E_Q}
  \end{center}
  \vspace*{-.225in}
\end{figure}

Using Table~\ref{tab:optimal} values and taking the approximations, 
\begin{eqnarray*}
\omega^{I}_{d} \approx \sqrt{\frac{1}{L_p C_{T,max}}}\mbox{ }\mbox{and}\mbox{ }\omega^{II}_{d} \approx \sqrt{\frac{1}{L_p C_{T,min}}},\\
Q^{II}_{f}=\frac{\omega^{II}_{d}L}{R_{L}}\approx\sqrt{\beta}Q^{I}_{f}\mbox{ }\mbox{and}\mbox{ }\ln{\alpha^{II}}\approx\frac{\ln{\alpha^{II}}}{\sqrt{\beta}}, 
\end{eqnarray*}
we study the effect of quality factor on the optimal gains of P-SSHI and S-SSHI circuits over FWR as in Fig.~\ref{E_Q}.

%---------------------------------------------------------------
\vspace*{-.05in}
\begin{figure*}[htbp]
\vspace*{-.1in}
  \begin{center}
  \includegraphics[width=1\linewidth]{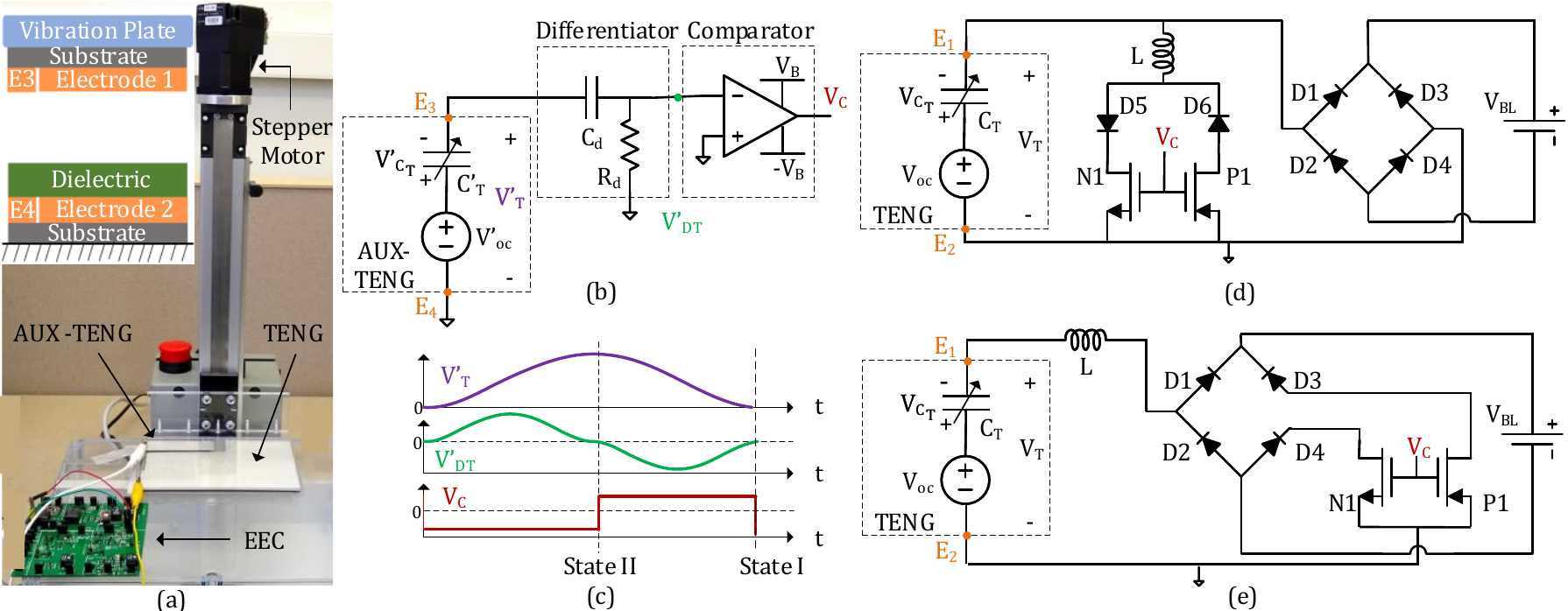}
  \vspace{-.25in}
  \caption{(a) Experimental Setup; Inset: Cross section view of a contact-separation TENG with Aux-TENG (b) Control circuit for P-SSHI and S-SSHI circuits (c) Control circuit waveforms (d) P-SSHI circuit implementation (e) S-SSHI circuit implementation}\label{StripI}
  \end{center}
\vspace*{-.35in}
\end{figure*}
%---
\begin{itemize}
\item The optimal gains of P-SSHI and S-SSHI circuit are increasing with inductor quality factor (Fig.~\ref{E_Q}(Top)).
\item For $\beta=1$, the optimum gains turn out to be the same for both P-SSHI and S-SSHI circuits, and it improves with increasing $\beta$ for both circuits (Fig.~\ref{E_Q}(Top)).
\item For any $\beta>1$, the optimal gain of P-SSHI is higher than S-SSHI (Fig.~\ref{E_Q}(Top)), but this is achieved for optimal battery load several times the maximum open circuit voltage ($V_{oc,max}$) depending on the quality factor (Fig.~\ref{E_Q}(Bottom)). TENGs typically have high $V_{oc,max}$ (in tens to hundreds of volts), and hence implementing a P-SSHI circuit at optimal load shall be challenging. 
\item In contrast, optimal load for S-SSHI circuit is independent of quality factor and lower than $V_{oc,max}$ (Fig.~\ref{E_Q}(Bottom)). 
\end{itemize}
Further, the effect of battery load on the energy output is studied in section \ref{Results}.
\vspace*{-.15in}
\section{Experimental Implementation}
\vspace*{-.05in}
\subsection{Experimental Setup}
\vspace*{-.05in}
TENG with a contact area of $113.8~cm^2$ is built as shown in the schematic of Fig.~\ref{model} using Al foil as the electrodes and Teflon of thickness $127~\mu m$ as the dielectric. The bottom plate is fixed while the top plate is driven in a vertical contact-separation mode using a programmed stepper motor at $10~Hz$ frequency and maximum separation $(x_{max})$ of $1.2~mm$. The setup picture is shown in Fig.~\ref{StripI}(a). 
\vspace*{-.2in}
\subsection{TENG Characterization}
\vspace*{-.05in}
The key electrical parameters to characterize TENG are maximum open circuit voltage ($V_{oc,max}$), minimum and maximum TENG capacitance ($C_{T,min}$ and $C_{T,max}$).
\subsubsection{TENG Capacitance}
We follow the method described in \cite{ghaffarinejad2018conditioning},\cite{lu2016batch} to measure the dynamic variation of the TENG Capacitance ($C_T(t)$) and extract the minimum and maximum value ($C_{T,min}$ and $C_{T,max}$). The details of the method are repeated in the Appendix~\ref{A_Cap} along with the plot of measured $C_T(t)$.
\subsubsection{TENG maximum open circuit voltage}
We use the standard FWR circuit to measure $V_{oc,max}$. In the section \ref{FWR_UB}, the upper bound on the load battery voltage is derived for the FWR circuit. Instead of a battery, if a capacitor (of any capacitance) is charged using a FWR circuit, its voltage will saturate at the same value ($V_{sat}$), which by (\ref{Eq:15}) satisfies:
\vspace*{-.1in}
\begin{equation}
    V_{oc,max} = (\beta+1)(V_{sat}+2V_D).\nonumber
    \vspace*{-.1in}
\end{equation}
By charging a capacitor, $V_{sat}$ was measured as $54.4V$. At saturation, negligible current flows through the FWR circuit, and hence diode voltage drop ($V_D$) can be neglected, yielding $V_{oc,max}$ as $188.77V$. The measured TENG parameters are summarized in Table~\ref{tab:Para}.
\begin{table}[ht]
\vspace*{-.025in}
    \centering
    \caption{Measured TENG Parameters}
    \label{tab:Para}
    \vspace*{-.1in}
    \begin{tabular}{c|c}
    \hline\\
    Maximum open circuit voltage: $V_{oc,max}$  & 188.77 V\\[1.2ex]
    
    Minimum TENG capacitance: $C_{T,min}$ & 97.28 pF\\[1.2ex]
    
    Maximum TENG capacitance: $C_{T,max}$ & 239.80 pF\\[1.2ex]
    
    TENG capacitance ratio: $\beta$ & 2.47\\[1.2ex]
    \hline
    \end{tabular}
    \vspace*{-0.25in}
\end{table}
\begin{figure*}[htb]
\vspace*{-.1in}
  \begin{center}
  \includegraphics[width=1\linewidth]{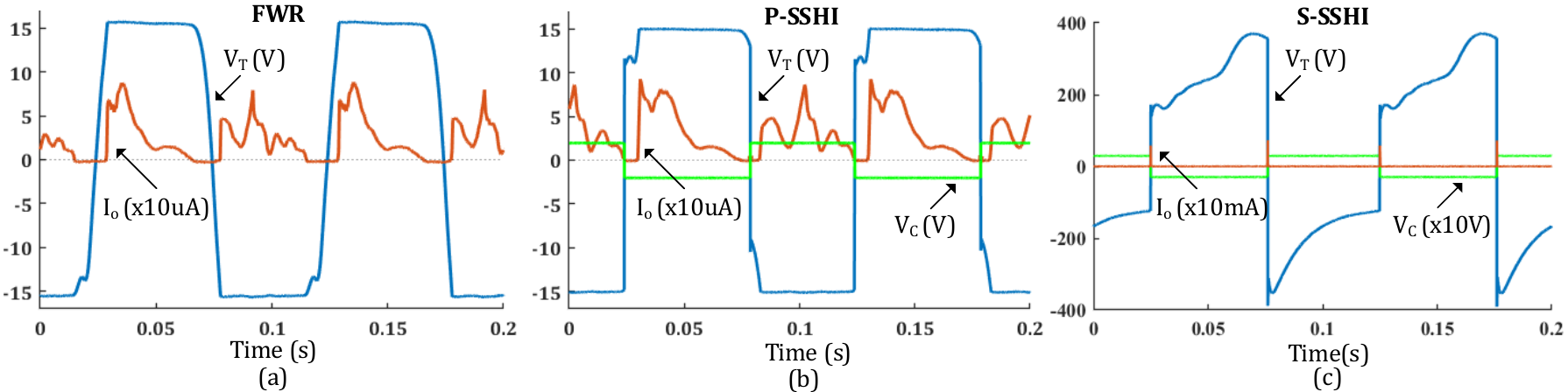}
  \vspace{-.25in}
  \caption{Measured TENG Voltage ($V_T$), load current ($I_o$) and (if) control voltage ($V_C$) waveforms at 15V battery load of the (a) FWR, (b) P-SSHI, and (c) S-SSHI circuits.}\label{W_all}
  \end{center}
\vspace*{-.5in}
\end{figure*}
\begin{figure} [hbtp]
  \vspace*{.15in}
  \begin{center}
  \includegraphics[width=1\linewidth]{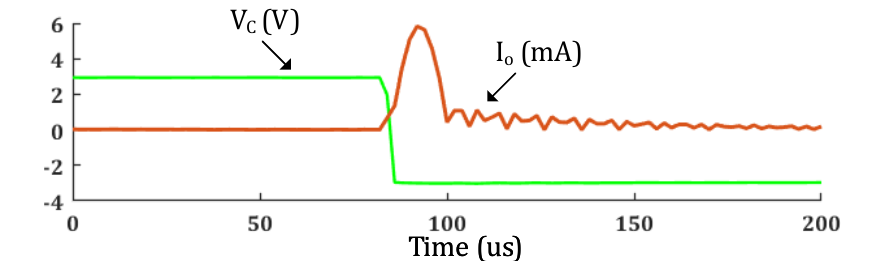}
  \vspace*{-.125in}
  \caption{Zoomed in measured load current ($I_o$) and control voltage ($V_C$) waveforms during switching at State I.}\label{SSHI_M_Zoom}
  \end{center}
  \vspace*{-.35in}
\end{figure}
\vspace*{-.15in}
\subsection{Implementation of Energy Extraction Circuits}
\vspace*{-.05in}
Synchronous switched circuits (P-SSHI and S-SSHI) described above are implemented using the MOSFET as switches and operated through dedicated control circuits. The control circuit's function is to detect States I and II during the TENG operation and issue the required gate pulse to the MOSFET switches. The switching control circuit receives input from an ``auxiliary" TENG (with contact area $1/8^{th}$ to that of the main TENG) that is implemented on the same fixed bottom and moving upper plates as the main TENG to operate in parallel (Fig.~\ref{StripI}(a)). This isolation of auxiliary TENG (from the main TENG) allows non-interfering operation of the control circuit independent of the TENG voltage waveform and prevents charge leakage (and subsequent performance degradation) from the main TENG into the control circuit. The energy consumed by the control circuit can be deduced separately and is given in Table \ref{tab:Control}. Off-the-shelf components used for the implementation are listed in Appendix~\ref{A_List}.
\begin{table}[ht]
\vspace*{-.125in}
    \begin{center}
    \caption{Control Circuit Power Consumption}
    \label{tab:Control}
    \vspace*{-.1in}
    \begin{tabular}{c|c|c|c}
    \hline
    Circuit & Avg. Current& Supply Voltage ($2V_B$) & Power \\
    \hline
    P-SSHI            & 257 nA                    & 4 V                        & 1.03 uW             \\
S-SSHI            & 590 nA                    & 7 V                        & 4.13 uW             \\
    \hline
    \end{tabular}
    \end{center}
    \vspace*{-0.1in}
\end{table}

\subsubsection{FWR Circuit}
Implementation of the FWR circuit that contains no synchronized switches is straightforward, as shown in Fig.~\ref{FWR}.
Fig.~\ref{W_all}(a) shows the measured TENG voltage ($V_T$) and load current ($I_o$) for the battery load ($V_{BL}$) of 15V.

\subsubsection{P-SSHI Circuit}
P-SSHI circuit is implemented as shown in Fig.~\ref{StripI}(d). Its synchronized switching control circuit consists of a RC differentiator followed by a zero-crossing comparator (Fig.~\ref{StripI}(b)). Value of RC differentiator is chosen to provide high output impedance to the Aux-TENG (``auxiliary" TENG); hence it operates in near open circuit condition ($V'_{T}$). The differentiated signal ($V'_{DT}$) is zero at the extrema: States I and II, which triggers the state change of the comparator ($V_C$) (Fig.~\ref{StripI}(c)). The NMOS switch is turned on at State II and PMOS at State I due to state change in gate signal ($V_C$). The switch turn-off is automatic since the current direction reversal at the end of half the $LC_T$ oscillation time-period is blocked by the corresponding diode $D5$ (or $D6$). The control circuit is powered by $\pm 2V$ external source. Fig.~\ref{W_all}(b) shows the measured TENG voltage ($V_T$), control signal ($V_C$), and load current ($I_o$) for the battery load ($V_{BL}$) of 15V. Note the instants of voltage inversion at States I and II compared to the FWR circuit TENG voltage (Fig.~\ref{W_all}(a)).

\subsubsection{S-SSHI Circuit}
Here, the same control circuit as used for the P-SSHI circuit is used (Fig.~\ref{StripI}(b)) and is powered by $\pm 3.5V$ supply. Fig.~\ref{StripI}(e) shows the implementation of the S-SSHI circuit. The circuit operation is the same as the P-SSHI circuit explained above, with the inductor, NMOS and PMOS switches in series connection with the TENG compared to parallel as in the case of the P-SSHI circuit. Fig.~\ref{W_all}(a) shows the measured steady-state TENG voltage ($V_T$), control signal ($V_C$), and load current ($I_o$) for the battery load $V_{BL}=15V$. Note the TENG voltage at State II reaches almost twice the maximum open circuit voltage of the TENG ($V_{oc,max}$) due to the transient buildup via voltage inversions. Fig.~\ref{SSHI_M_Zoom} shows the zoomed in view of the $I_o$ and $V_C$ during the switching at State I. The load current ($I_o$) is a half sinusoid since the switch is on for half the $LC_T$ oscillation time period. The post-switching ringing transients observed in $I_o$ can be attributed to the stray parasitics of the circuit.
\begin{figure*}[htb]
\vspace*{-.1in}
  \begin{center}
  \includegraphics[width=1\linewidth]{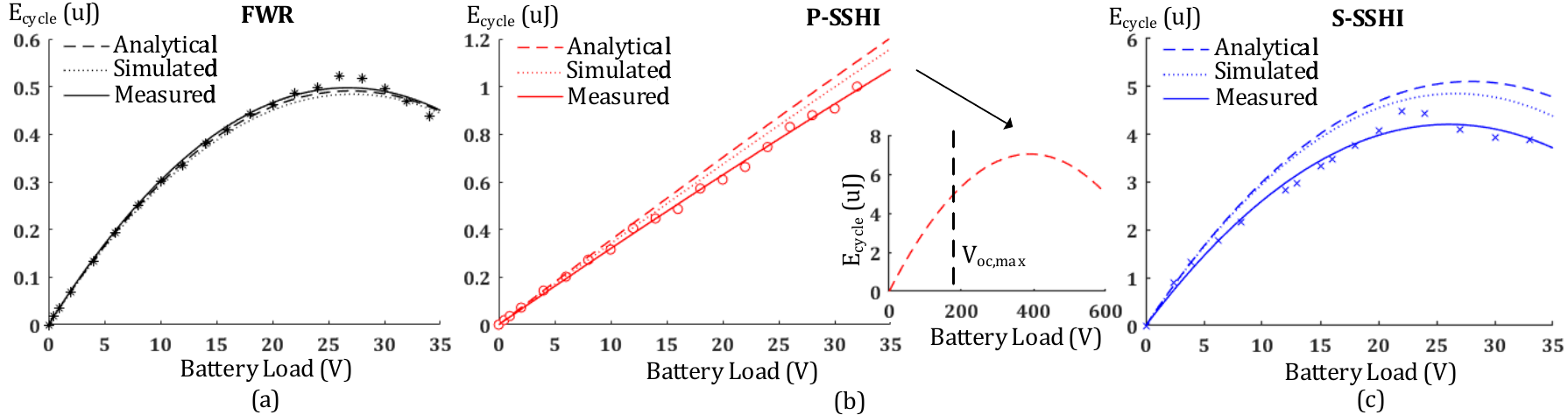}
  \vspace{-.25in}
  \caption{Analytical, simulated and measured per-cycle energy plot for (a) FWR, (b) P-SSHI; Inset: Extrapolated P-SSHI analytical plot, and (c) S-SSHI circuits against the battery load.}\label{E_all}
  \end{center}
\vspace*{-.3in}
\end{figure*}
\begin{figure}[htb]
  \begin{center}
  \includegraphics[width=1\linewidth]{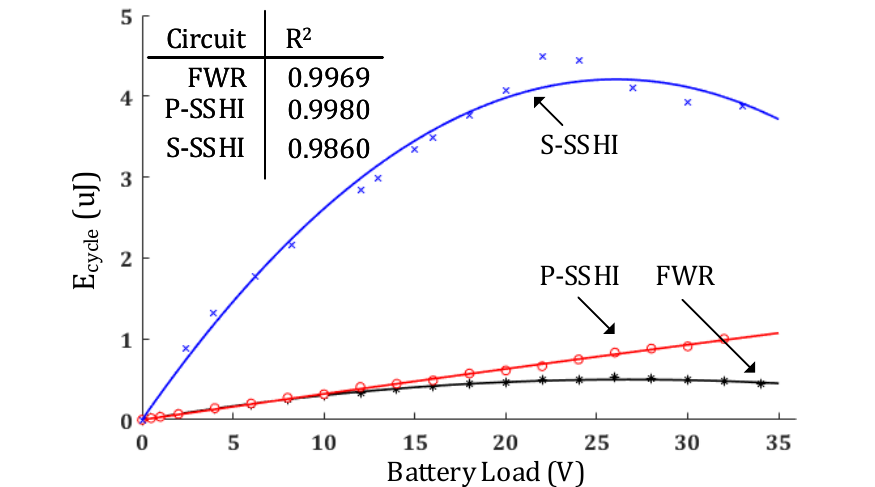}
  \vspace{-.2in}
  \caption{Comparison of measured per-cycle energy output for FWR, P-SSHI, and S-SSHI circuit against the battery load.}\label{E_M_all}
  \end{center}
\vspace*{-.35in}
\end{figure}
\vspace*{-.15in}
\section{Energy Results and Discussion}\label{Results}
\vspace*{-.05in}
The experimentally measured, simulated, and analytically obtained per-cycle energy output ($E_{cycle}$) at different load battery values ($V_{BL}$) for each of the three considered energy extraction circuits (EECs) are co-plotted in Fig.~\ref{E_all}. Experimental results were obtained using the implemented TENG with parameters listed in Table~\ref{tab:Para}. For simulation, the TENG circuit model described in Sec.~\ref{TENG_Circuit_Model} was implemented in a PSpice tool using the same measured parameters of Table~\ref{tab:Para} and operation frequency of $10~Hz$. The circuits were built using the PSpice models of the off-the-shelf components (diodes, MOSFETs, and comparator) used in experimental implementation available from the vendor websites. The commercial inductor used in the P-SSHI/S-SSHI circuits was modeled as a series $L$-$R_L$ for simulation with frequency dependent inductance and resistance values obtained experimentally using the RLC meter at the respective $L$-$C_T$ resonance frequencies of States I and II: $\omega^{I}_{d}$ and $\omega^{II}_{d}$ (refer Appendix~\ref{A_Ind_C} for details). While for the analytical results, measured TENG parameters (Table~\ref{tab:Para}) and the non-ideality parameters estimated from the manufacturer provided PSpice models and the simulation of those models at multiple operating points were used in the equations derived in Sec.~\ref{EEC}. The parameter values of the considered non-idealities: Diode voltage drop ($V_D$), Leakage charges ($Q_L^I$ and $Q_L^{II}$), and the normalized quality factors ($\alpha^{I}$ and $\alpha^{II}$), that change significantly over the operating range of the battery load, are tabulated for each circuit in Appendix~\ref{A_N_Para} Tables~\ref{tab:FWR_Para}-\ref{tab:SSHI_Para} and were estimated at each 2V increment of battery load using their PSpice model simulations.
\vspace*{-.15in}
\subsection{FWR Circuit}
\vspace*{-.05in}
Referring to Fig.~\ref{E_all}(a), $E_{cycle}$ shows the parabolic behavior to the change in $V_{BL}$ with peak at $26.7V$, as expected from the analytical derivation of (\ref{eq:F4}). A close match between the analytical, simulated, and measured values is achieved. 
\vspace*{-.15in}
\subsection{P-SSHI Circuit}
\vspace*{-.05in}
The $E_{cycle}$ for the P-SSHI circuit strictly increases with $V_{BL}$ in the considered range of $0-35 V$ (Fig.~\ref{E_all}(b)). As expected from the analysis (\ref{eq:P13}), the response is parabolic in nature with a peak obtained at battery load several times the maximum open circuit voltage ($V_{oc,max}$). Extrapolation of the analytical curve (inset Fig.~\ref{E_all}(b)) indeed confirms the same with the peak around $400V$, keeping in mind that the extrapolation is based on the used non-ideality parameters at lower operating voltages.
\vspace*{-.15in}
\subsection{S-SSHI Circuit}
\vspace*{-.05in}
For the S-SSHI circuit, the parabolic response gives the peak $E_{cycle}$ at around $26V$ (Fig.~\ref{E_all}(c)). For an ideal circuit, the peak is expected at the same voltage as the FWR circuit (refer Table \ref{tab:optimal}), but the leakage and diode drop shifts the peak lower. S-SSHI circuit has a high steady-state voltage of almost $400V$ (Fig.~\ref{W_all}(c)), which requires the use of high-rated MOSFETs (refer Appendix~\ref{A_List} for details) accompanied with significant leakage charge (refer cols 5 and 6 in Table~\ref{tab:SSHI_Para} of Appendix~\ref{A_N_Para}). Hence, leakage charge becomes the dominant non-ideality affecting the S-SSHI circuit's performance. Also, stray parasitics such as inductor and MOSFET's parallel capacitance give rise to post-switching ringing transients as can be seen in the measured circuit current (Fig.~\ref{SSHI_M_Zoom}). Such ringing leads to a post-switching additional loss of TENG charge, and so an immediate decrease (in absolute terms) in the post switching voltage (refer Fig.~\ref{W_all}(c)), resulting in reduced $E_{cycle}$. This additional loss in the TENG charge is also captured in the analytical results through the added leakage charge term $Q_L$. Any additional remaining discrepancy between the experimental and the analytical results may be attributed to the unaccounted factors such as practical delay in the control circuit that leads to switching past the extrema (States I and II); measurement errors in the TENG parameters of Table~\ref{tab:Para}; the TENG's own parasitic resistance (due to imperfect connection to the electrode, connection wires); etc.

\vspace*{-.2in}
\subsection{Overall Comparison}
\vspace*{-.05in}
Fig.~\ref{E_M_all} plots the measured $E_{cycle}$ for all the considered circuits. At lower battery loads, the P-SSHI circuit has output almost equal to the FWR circuit as expected: The MOSFET switches fail to turn ``on" since the battery load equals the drain to source voltage ($V_{ds}$) of the MOSFETs, that turns the P-SSHI circuit of Fig.~\ref{StripI}(d) to simply the FWR circuit of Fig.~\ref{FWR}. The S-SSHI circuit has the highest output in the considered load range. The two switched inductor circuits require overhead power for the control circuit (Table~\ref{tab:Control}), meaning their net output and gain over FWR is lower. For example, in our implementation, the P-SSHI control circuit requires $1.03~\mu W$. Hence, the operation at $10 Hz$ extracts net positive energy for battery load greater than $3.2V$ and has increased gain over FWR circuit post battery load of $16.75V$. 
%--------------------------------------
\vspace*{-.175in}
\section{Conclusion}
\vspace*{-.05in}
This work developed a mathematical analysis framework for the synchronous switched energy extraction circuits: P-SSHI and S-SSHI  along with the standard FWR circuit for TENG transducers and derived closed-form formulae for their per-cycle energy output, optimum battery load value, and also the upper bound on battery voltage beyond which extraction is not feasible. The modeling included the non-idealities of diode drop, leakage current, and switch and inductor resistances. The strong match of analytical model results with simulation and measured ones shows that the analytical models can be used to assess TENG EEC performance once the TENG parameters have been measured. 

An ideally switched P-SSHI/S-SSHI circuit at optimal load was found to provide more than 100 fold gain over the FWR circuit at their optimum values, depending on the inductor quality factor. The effect of load battery change on the per-cycle energy output was examined. All the circuits show a parabolic response, providing an optimal battery load and an upper bound. 
 
Such comparison of energy extraction circuits brought forward their pros versus cons over each other. S-SSHI has superior energy output compared to other circuits but operates over a smaller load range. While P-SSHI energy output can exceed that of S-SSHI, it is optimized at a load voltage value several times the TENG maximum open circuit voltage ($V_{oc,max}$). Use of a DC/DC converter can make this feasible, but with added complexity and loss, and so P-SSHI shall be more suited for TENG with low $V_{oc,max}$. Also, for a TENG with low $V_{oc,max}$ or high $\beta$, the low upper bound on the battery load in the case of the S-SSHI circuit can be limiting. In summary, the presented work offers design insights/tradeoffs for TENG energy extraction circuits.

A possible direction of further research is to implement the discussed circuit architectures by using ``passive" control circuits. Development of integrated efficient DC/DC converters to operate the P-SSHI and S-SSHI circuit at their optimal load for any given TENG and the on-board battery is another direction. Further, implementing the EECs identified leakage through MOSFET switches as one of the major loss factors. So other switching technologies with high rated voltage and low leakage, such as SiC (Silicon Carbide), can be explored.  
%--------------------------------

\vspace*{-.15in}
\appendix
\subsection{Dynamic measurement of TENG capacitance}\label{A_Cap}
\begin{figure} [htbp]
  \vspace*{-.1in}
  %\hspace*{-.2in}
   \includegraphics[width=1\linewidth]{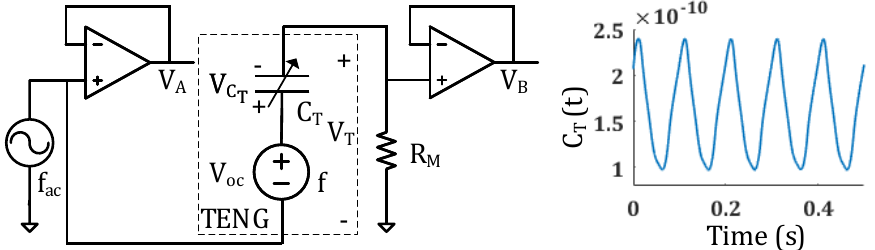}
  \vspace*{-.1in}
  \caption{Measurement setup for TENG capacitance ($C_T(t)$) and its plot}\label{fig:captest}
  \vspace*{-.2in}
\end{figure}

\vspace*{-.05in}
An AC voltage source with frequency $f_{ac}$ several magnitudes higher than the TENG operation frequency $f$ was connected to TENG (see Fig.~\ref{fig:captest}). 
Using a high input impedance voltage follower op-amp, Voltages V$_A$ and V$_B$ were observed on the two channels of oscilloscope. With V$_A$ as reference, it can be easily showed that, $C_T=\frac{1}{2\pi f_{ac}R_Mtan(\theta)}$,
where $\theta$ is the phase difference between the AC voltages V$_A$ and V$_B$. Fig.~\ref{fig:captest} plot shows the measured $C_T(t)$, from which we determined, $C_{T,max}=239.80 pF$ and $C_{T,min}=97.28 pF$.

%---
\vspace*{-.15in}
\subsection{List of off-the-shelf components used}\label{A_List}
\vspace*{-.05in}
All MOSFETs were sourced from Diodes Inc. (diodes.com), inductor from Bourns Inc. (bourns.com), didoes from ON Semiconductors (onsemi.com), and TS881 used as comparator in the control circuit from STMicroelctronics (st.com). PSpice models used were downloaded from vendor websites.
\begin{table}[ht]
 \vspace*{-0.1in}
    \centering
    \caption{List of discrete components used}\label{tab:list}
     \vspace*{-0.1in}
    \begin{tabular}{c|c|c|c}
    \hline
Circuit & FWR    & P-SSHI        & S-SSHI               \\
\hline
Diode & 1N4148   & 1N4148    & MUR160       \\
NMOS  & -   & DMN601K     & DMN60H080DS     \\
PMOS  & -   & DMP510DL  & DMP45H150DE  \\
Inductor    &-   &RLB1014-104KL   &RLB1014-104KL \\
\hline
    \end{tabular}
    \vspace*{-.2in}
\end{table}
\vspace*{-.1in}
\subsection{Inductor Characterization}\label{A_Ind_C}
\vspace*{-.05in}
The inductor is modeled as series $L$-$R_L$ for the simulation. Both $L$ and $R_L$ were measured using the RLC meter at the State I and State II resonance frequency ($\omega_d^{I}$ and $\omega_d^{I}$) and are tabulated below. The nominal inductance for RLB1014-104KL used in P-SSHI/S-SSHI circuit  was $100mH$.
\begin{table}[ht]
\vspace*{-0.1in}
    \centering
    \caption{Inductor Characterization}\label{tab:Ind}
    \vspace*{-0.1in}
    \begin{tabular}{c|c|c}
    \hline
Frequency        &$L(mH)$ &$R_L(\Omega)$ \\
    \hline
$\omega_d^{I}=$227.02 kHz & 80.91   &296    \\[1.2ex]
$\omega_d^{II}=$343.55 kHz & 87.09  &395.6   \\ [1.2ex]        
    \hline
    \end{tabular}
    \vspace*{-0.2in}
\end{table}
\vspace*{-.15in}
\subsection{Non-ideality parameters}\label{A_N_Para}
\vspace*{-.05in}
Table~\ref{tab:FWR_Para} lists the values of the diode voltage drops ($V_D$), averaged across all the diodes, for the FWR circuit, and was obtained from manufacturer PSpice models and their simulation at the increments of 2V battery load ($V_{BL}$).

\begin{table}[h]
\vspace*{-.15in}
\centering
\caption{FWR circuit's average diode voltage drop at different battery load}\label{tab:FWR_Para}
    \vspace*{-0.1in}
\begin{tabular}{c|c|c|c} 
\hline
$V_{BL}(V)$  & $V_D(mV)$       & $V_{BL}(V)$  & $V_D(mV)$        \\ 
\hline
2  & 239.190  & 24 & 239.000     \\
4  & 239.195 & 26 & 235.895  \\
6  & 239.170  & 28 & 236.880   \\
8  & 240.010  & 30 & 233.820   \\
10 & 241.085 & 32 & 232.110   \\
12 & 241.715 & 34 & 231.190   \\
14 & 240.405 & 36 & 229.975  \\
16 & 240.890  & 38 & 228.430   \\
18 & 239.935 & 40 & 225.680   \\
20 & 239.660  & 42 & 222.920   \\
22 & 238.615 & 44 & 219.840  \\
\hline
\end{tabular}
\end{table}

Table~\ref{tab:PSSHI_Para} and Table~\ref{tab:SSHI_Para} list the values of non-ideality parameters obtained from manufacturer PSpice models and the simulation of those models at increments of 2V battery load ($V_{BL}$) for the P-SSHI and S-SSHI circuits, respectively. In both the tables, the diode voltage drop: $V_D$ averaged across all diodes is listed first. Next two columns list $\alpha^{I}$ and $\alpha^{II}$,  the normalized resonator quality factor during switching at States I and II, respectively. In the last two columns, $Q_L^I$ and $Q_L^{II}$ list the leakage charge during the first half-cycle and second half-cycle, respectively.
\begin{table}[h]
\centering
\caption{P-SSHI circuit's non-ideality parameters at different battery load}\label{tab:PSSHI_Para}
    \vspace*{-0.1in}
\begin{tabular}{c|c|c|c|c|c}
\hline
$V_{BL}(V)$  & $V_D(mV)$       & $\alpha^{I}$           & $\alpha^{II}$           & $Q_L^I(nC)$           & $Q_L^{II}(nC)$            \\
\hline
2  & 238.630 & 0.982685 & 0.992048 & 0.417454 & 0.336543  \\
4  & 238.610 & 0.964918 & 0.972427 & 0.471214 & 0.355721  \\
6  & 238.580 & 0.962172 & 0.963049 & 0.527053 & 0.374757  \\
8  & 238.530 & 0.958906 & 0.962619 & 0.573477 & 0.392414  \\
10 & 238.480 & 0.956383 & 0.959136 & 0.618310  & 0.410486  \\
12 & 238.360 & 0.955017 & 0.960197 & 0.663186 & 0.429357  \\
14 & 238.565 & 0.954870 & 0.961392 & 0.705960  & 0.447556  \\
16 & 238.485 & 0.954431 & 0.960071 & 0.748835 & 0.463513  \\
18 & 238.375 & 0.954023 & 0.959017 & 0.787447 & 0.480360  \\
20 & 238.260 & 0.953624 & 0.958170 & 0.825341 & 0.497374  \\
22 & 238.115 & 0.953273 & 0.958331 & 0.866459 & 0.516530  \\
24 & 237.955 & 0.952561 & 0.958361 & 0.906872 & 0.531336  \\
26 & 237.780 & 0.952291 & 0.957609 & 0.937981 & 0.549390  \\
28 & 237.620 & 0.952431 & 0.957765 & 0.979318 & 0.565194  \\
30 & 237.420 & 0.951825 & 0.958160 & 1.012891 & 0.582427  \\
32 & 237.210 & 0.952550 & 0.957170 & 1.046312 & 0.598388  \\
34 & 236.970 & 0.951946 & 0.957555 & 1.075120 & 0.616600  \\
36 & 236.730 & 0.951743 & 0.957006 & 1.119814 & 0.631156  \\
38 & 236.480 & 0.951257 & 0.957007 & 1.169261 & 0.647248  \\
40 & 236.195 & 0.951298 & 0.956769 & 1.216217 & 0.666166  \\
42 & 235.915 & 0.951122 & 0.957339 & 1.270956 & 0.689827  \\
44 & 235.595 & 0.951433 & 0.956825 & 1.360819 & 0.712828  \\
\hline
\end{tabular}
\vspace*{-0.1in}
\end{table}

\begin{table}[h]
\centering
\caption{S-SSHI circuit's non-ideality parameters at different battery load}\label{tab:SSHI_Para}
    \vspace*{-0.1in}
\begin{tabular}{c|c|c|c|c|c}
\hline
$V_{BL}(V)$  & $V_D(mV)$       & $\alpha^{I}$           & $\alpha^{II}$           & $Q_L^I(nC)$           & $Q_L^{II}(nC)$            \\
\hline
2  & 711.620  & 0.824141 & 0.973194 & 6.729874 & 16.094671  \\
4  & 711.205 & 0.827699 & 0.973968 & 6.644104 & 15.083334  \\
6  & 710.860  & 0.829109 & 0.973642 & 6.570467 & 13.609666  \\
8  & 710.585 & 0.827362 & 0.97366  & 6.515866 & 12.701908  \\
10 & 710.065 & 0.831173 & 0.971093 & 6.434156 & 11.453035  \\
12 & 709.680  & 0.833499 & 0.969869 & 6.402619 & 10.413612  \\
14 & 708.965 & 0.838585 & 0.962589 & 6.410849 & 9.182698   \\
16 & 707.990  & 0.844684 & 0.964754 & 6.421081 & 8.230673   \\
18 & 707.440  & 0.851579 & 0.961934 & 6.453288 & 7.315406   \\
20 & 701.105 & 0.861117 & 0.961399 & 6.474542 & 6.541747   \\
22 & 705.045 & 0.872938 & 0.95959  & 6.571256 & 5.896586   \\
24 & 703.925 & 0.881601 & 0.958869 & 6.505392 & 5.273539   \\
26 & 703.040  & 0.895017 & 0.960909 & 6.471428 & 4.810095   \\
28 & 700.830  & 0.908616 & 0.954769 & 6.471593 & 4.286924   \\
30 & 699.540  & 0.924828 & 0.957486 & 6.472602 & 3.974769   \\
32 & 697.450  & 0.945902 & 0.957086 & 6.455887 & 3.526472   \\
34 & 692.645 & 0.960356 & 0.956701 & 6.499851 & 2.90527    \\
36 & 682.280  & 0.94225  & 0.94882  & 6.101366 & 1.892802   \\
38 & 673.480  & 0.943536 & 0.953631 & 5.625206 & 1.738115   \\
40 & 660.885 & 0.940743 & 0.948493 & 5.171155 & 1.474435   \\
42 & 647.110  & 0.940695 & 0.951682 & 4.715545 & 1.242598   \\
44 & 629.710  & 0.942761 & 0.950688 & 4.058914 & 1.021044   \\
\hline
\end{tabular}
\vspace*{-.25in}
\end{table}
%-------------------
\vspace*{-.05in}
\bibliographystyle{IEEEtran}
\bibliography{IEEEabrv,Bibliography}
%--------

% or


% ============================================
%\appendices

%\section{}

%\section*{Acknowledgment}

\ifCLASSOPTIONcaptionsoff
  \newpage
\fi

\vfill

\end{document}